\documentclass[twocolumn]{aastex631}

\hypersetup{linkcolor=cyan,citecolor=cyan,filecolor=cyan,urlcolor=cyan}

\accepted{for publication in AJ}

\shorttitle{ACCESS: HATS-5b}
\shortauthors{Allen et al.}
\graphicspath{{./}{figures/}}

\begin{document}

\title{ACCESS: Tentative detection of H$_2$O in the ground-based optical transmission spectrum of the low-density hot Saturn HATS-5b}

\author[0000-0002-0832-710X]{Natalie H. Allen}\altaffiliation{NSF Graduate Research Fellow}
\affiliation{William H. Miller III Department of Physics and Astronomy, Johns Hopkins University, Baltimore, MD 21218, USA}

\author[0000-0001-9513-1449]{N\'estor Espinoza}
\affiliation{Space Telescope Science Institute, 3700 San Martin Drive, Baltimore, MD 21218, USA}
\affiliation{William H. Miller III Department of Physics and Astronomy, Johns Hopkins University, Baltimore, MD 21218, USA}

\author[0000-0002-5389-3944]{Andr\'es Jord\'an}
\affiliation{Facultad de Ingenier\'ia y Ciencias, Universidad Adolfo Ib\'a\~nez, Av.\ Diagonal las Torres 2640,
Pe\~nalol\'en, Santiago, Chile}
\affiliation{Millennium Institute for Astrophysics, Chile}
\affiliation{Data Observatory Foundation, Chile}

\author[0000-0003-3204-8183]{Mercedes L\'opez-Morales}
\affiliation{Center for Astrophysics ${\rm \mid}$ Harvard {\rm \&} Smithsonian, 60 Garden St, Cambridge, MA 02138, USA}

\author[0000-0003-3714-5855]{D\'aniel Apai}
\affiliation{Steward Observatory, The University of Arizona, 933 N. Cherry Avenue, Tucson, AZ 85721, USA}
\affiliation{Lunar and Planetary Laboratory, The University of Arizona, 1629 E. University Boulevard, Tucson, AZ 85721}

\author[0000-0002-3627-1676]{Benjamin V.\ Rackham}
\altaffiliation{51 Pegasi b Fellow}
\affiliation{Department of Earth, Atmospheric and Planetary Sciences, and Kavli Institute for Astrophysics and Space
Research, Massachusetts Institute of Technology, Cambridge, MA 02139, USA}

\author[0000-0002-4207-6615]{James Kirk}
\affiliation{Center for Astrophysics ${\rm \mid}$ Harvard {\rm \&} Smithsonian, 60 Garden St, Cambridge, MA 02138, USA}

\author[0000-0003-0412-9664]{David J. Osip}
\affiliation{Las Campanas Observatory, Carnegie Institution for Science, Colina el Pino, Casilla 601 La Serena, Chile}

\author[0000-0001-6205-6315]{Ian C. Weaver}
\affiliation{Center for Astrophysics ${\rm \mid}$ Harvard {\rm \&} Smithsonian, 60 Garden St, Cambridge, MA 02138, USA}

\author[0000-0002-6167-3159]{Chima McGruder}\altaffiliation{NSF Graduate Research Fellow}
\affiliation{Center for Astrophysics ${\rm \mid}$ Harvard {\rm \&} Smithsonian, 60 Garden St, Cambridge, MA 02138, USA}

\author[0000-0003-3455-8814]{Kevin Ortiz Ceballos}\altaffiliation{NSF Graduate Research Fellow}
\affiliation{Center for Astrophysics ${\rm \mid}$ Harvard {\rm \&} Smithsonian, 60 Garden St, Cambridge, MA 02138, USA}

\author[0000-0001-6533-6179]{Henrique Reggiani}
\altaffiliation{Carnegie Fellow}
\affiliation{The Observatories of the Carnegie Institution for Science, 813 Santa Barbara St, Pasadena, CA 91101, USA}

\author[0000-0002-9158-7315]{Rafael Brahm}
\affiliation{Facultad de Ingenier\'ia y Ciencias, Universidad Adolfo Ib\'a\~nez, Av.\ Diagonal las Torres 2640,
Pe\~nalol\'en, Santiago, Chile}
\affiliation{Millennium Institute for Astrophysics, Chile}

\author[0000-0003-0650-5723]{Florian Rodler}
\affiliation{European Southern Observatory, Alonso de Cordova 3107, Vitacura, Santiago de Chile, Chile}

\author[0000-0002-8507-1304]{Nikole K Lewis} \affiliation{Department of Astronomy and Carl Sagan Institute, Cornell
University \\ 122 Sciences Drive, Ithaca, NY, 14853, USA}

\author[0000-0003-0910-5805]{Jonathan Fraine}
\affiliation{Space Science Institute Center for Data Science, USA}
\affiliation{Space Science Institute Center for Exoplanet and Planetary Science, USA}



\begin{abstract}
We present a precise ground-based optical transmission spectrum of the hot-Saturn HATS-5b ($T_{eq} =1025$ K), obtained as part of the ACCESS survey with the IMACS multi-object spectrograph mounted on the Magellan/Baade Telescope. Our spectra cover the 0.5 to 0.9 micron region, and are the product of 5 individual transits observed between 2014 and 2018. We introduce the usage of additional second-order light in our analyses which allows us to extract an “extra” transit light curve, improving the overall precision of our combined transit spectrum. We find that the favored atmospheric model for this transmission spectrum is a solar-metallicity atmosphere with sub-solar C/O, whose features are dominated by H$_2$O and with a depleted abundance of Na and K. If confirmed, this would point to a ``clear” atmosphere at the pressure levels probed by transmission spectroscopy for HATS-5b. Our best-fit atmospheric model predicts a rich near-IR spectrum, which makes this exoplanet an excellent target for future follow-up observations with the James Webb Space Telescope, both to confirm this H$_2$O detection and to superbly constrain the atmosphere's parameters. 
\end{abstract}



\section{Introduction} \label{sec:intro}
As we now know of thousands of planets orbiting stars other than the Sun, we have moved from the era of pure exoplanet detection into the realm of detailed planetary characterization. One of the most promising and exciting avenues for further discovery is exoplanet atmospheres. The atmospheres of exoplanets contain a wealth of information regarding many different scientific questions, from planet formation and orbital evolution \citep[see e.g.][]{Oberg:2011, Mordasini:2016, Madhu:2017, Espinoza:2017,  Reggiani_2022}, to signatures of planetary surfaces \citep[e.g.][]{Yu_2021}, to the potential presence of life revealed by biosignatures \citep[for a recent review, see, e.g., ][]{Schwieterman_2018}. 

The dominant method to study exoplanetary atmospheres currently is transmission spectroscopy. This technique has been used at both low- and high- resolution to discover a myriad of atomic and molecular features, from the cosmically abundant H$_2$O in the near-infrared, to molecules like TiO and VO and elements such as K and Na in the optical \citep[for recent reviews and discussion regarding the detections of species in exoplanet atmospheres, see e.g.][]{Deming_2017, Kreidberg_2018, Madhusudhan_2019}. These discoveries, in turn, have allowed us to start making inferences about the environments in which these atoms and molecules are present, including inferences about the presence of clouds and hazes in these distant worlds \citep[see, e.g.,][]{Sing:2016, Yu:2021}.

Ground-based observatories have played a significant role in our understanding of exoplanet atmospheres. Various ground-based observatories with multi-object spectrograph instruments, such as the Very Large Telescope (VLT) at Cerro Paranal Observatory with the FOcal Reducer/low dispersion Spectrograph 2 (FORS2) multi-object spectrograph \citep[see, e.g., ][and references therein]{Bean_2010, Nikolov_2018, Sedaghati_2017}, the Magellan Baade Telescope at Las Campanas Observatory with the Inamori-Magellan Areal Camera \& Spectrograph (IMACS) multi-object spectrograph \citep[see, e.g., ][and references therein]{Jord_n_2013, Rackham_2017, May_2019, Espinoza_2019}, the Large Binocular Telescope (LBT) at Mount Graham with the MODS instruments \citep[see, e.g.][]{Yan_2020}, and the Gemini Multi-Object Spectrograph (GMOS) instrument on the Gemini North/South telescope \citep[see, e.g., ][and references therein]{Gibson_2012, Todorov_2019}, among others, have been able to achieve transmission spectra in the optical wavelength range of comparable quality to those obtained by the \emph{Hubble Space Telescope} (\emph{HST}). This wavelength region is crucially important to constrain the presence of a wide variety of spectral features, from clouds and hazes \citep[see, e.g.,][]{LP:2016} to specific elements like Na and K \citep[see, e.g.,][]{Nikolov_2018} or molecules like TiO and VO \citep[see, e.g.,][]{Sedaghati_2017, Espinoza_2019, Sed:2021}. On top of this, given ground-based observatories are able to obtain the entire transit event, observations from high-precision ground-based facilities hold the promise to be able to extract features that are only observable during ingress/egress, such as light curve inhomogeneities due to morning-to-evening terminator variations \citep[see, e.g.,][]{Powell, Jones, espinoza_2021}. Chronicling and understanding exoplanetary atmospheres in the optical wavelength range will become even more crucial in the near-future in order to properly interpret results in the infrared from the \emph{James Webb Space Telescope} (\emph{JWST}).

HATS-5b is a short-period ($P = 4.76$ d), Saturn-sized gas giant ($R_p =0.868 \,\, R_J$, $M_p = 0.237 \,\, M_J$) orbiting a typical G-type star (R = $0.84 \,\, R_\odot$, M = $0.87 \,\, M_\odot$, $T_\mathrm{eff} = 5315$ K, $[Fe\slash H] = 0.04$) \citep{Zhou_2014}. With an equilibrium temperature of $T_{\rm eq} = 1025 \pm 17$ K, it sits on the very interesting limiting equilibrium temperature of 1000 K dividing ``hot" and ``warm" Jupiters \citep{Thorngren2016}. This regime is not only thought to divide exoplanets whose relatively large radii is impacted by external stellar irradiation \citep{Thorngren2016}, but also span a temperature range in which studies have predicted clearer atmospheres (and hence more feature-rich in transmission) than their hotter and cooler counterparts. At hotter temperatures, silicates are gaseous and thus form high-altitude clouds that wipe out atomic and elemental features in the transmission spectrum. At lower temperatures, photochemically produced hydrocarbon hazes mute atmospheric features to similar effect. These various haze and cloud effects have caused many exoplanet observations, especially at the depths probed in the optical, to appear flat and featureless. However, just at around 1000 K there is a potentially clear window between these two regions \citep[for a recent review of exoplanetary clouds and hazes and their effect on atmospheres, see][]{Gao_2021}. Planets with clear atmospheres make it easier to detect atomic and molecular features, which is key to infer details about the composition and formation histories of these distant worlds. In this work, we present a precise optical transmission spectrum of this low-density hot-Saturn HATS-5b, using the IMACS spectrograph \citep{Dressler2011} mounted on the Magellan Baade telescope. 

This paper is organized as follows. In \autoref{sec:observations} we describe the observations used for this study, including photometric monitoring of the host star. In \autoref{sec:reduction} we outline the data reduction and analysis used to extract the observables used to analyze both the Magellan/IMACS 
light curves and the resulting transit spectrum. In \autoref{sec:lcs}, we detail our light curve fitting and detrending approach, for both the white-light and wavelength-binned light curves, which were used to extract the transmission 
spectrum from our Magellan/IMACS data. In \autoref{sec:models} we interpret 
this transmission spectrum using exoplanet atmospheric modeling. We finally conclude with a summary of our findings in \autoref{sec:conclusions}.  

\section{Observations} \label{sec:observations}
\subsection{IMACS/Magellan}
We observed HATS-5b over a period of 4 years, from 2014 to 2018, with the IMACS instrument on the Magellan-Baade 6.5-meter telescope at Las Campanas Observatory as part of the ACCESS project\footnote{Originally the Arizona-CfA-Cátolica Exoplanet Spectroscopy Survey, since renamed to the Atmospheric Characterization Collaboration for Exoplanet Spectroscopic Studies.} \citep[][]{Rackham_2017}. We observed seven transits over seven nights in total, but we found two of these transits to be poorly affected by weather after reducing them and so we excluded them from the final analysis. The five transits analyzed in this work were obtained on 29 Oct 2014, 17 Dec 2015, 30 Nov 2017, 20 Dec 2017, and 27 Nov 2018, which we refer to in order as T1--T5, respectively. In addition, we were able to extract one extra transit light curve by making use of the second-order spectrum from the 2015 observing run, as the setup of the observation allowed for this light to be captured without contamination. We refer to this hereafter as T2.2. 

For all observations, we used the f/2 camera in the multi-object spectrograph configuration. IMACS has 8 chips, arranged in a $4 \times 2$ configuration, and in this f/2 mode the light is typically split across both vertically aligned chips. This camera has a very large field of view ($27.4'$), which, combined with the ability for custom slit masks, allows for simultaneous observations of multiple comparison stars in addition to our target star. For each observation of HATS-5, between five and seven comparison stars were also observed in order to track various systematics---including atmospheric conditions---during data collection. These stars are all of similar spectral type and are relatively close-by on sky; we list them and their properties in \autoref{tab:comp}. All seven comparison stars were used for T1 and T2, although the second order for comp2 fell off the chip and so could not be used for T2.2. For T3--T5, only comp1, comp2, comp4, comp8, and comp9 were observed, and all five of these comparison stars were used for T3 and T4. For T5, comp4 showed odd behavior in the white light curve and so was left out of the analysis. The custom slit masks were designed to be wide enough to minimize the potential for slit losses at typical seeing conditions at the observatory, while also being narrow enough to prevent contamination from nearby stars, or any more than necessary telluric signals. We use $10''$ wide and $22''$ long slits for T1, $10''$ wide and $10''$ long for T2 and T2.2, and $20''$ wide and $40''$ long slits for the remainder. The move to wider slits was made to see if obtaining more sky background would allow for a more complete removal of background noise, but no difference is seen in the final product's result between the two setups in our data. The lengths are chosen to cover enough sky past the spectra to properly subtract the sky background. These slit widths and lengths are used for all of the stars in the given setup.

\begin{deluxetable*}{c|| c c c c c c c c}
    \centering
    \tablecaption{A list of the various comparison stars used in this study. The 2MASS ID, RA and Dec are given for all targets, along with some magnitudes. $J$, $H$, and $K$ magnitudes are from 2MASS \citep{Skrutskie2006}, and $B$ and $V$ magnitudes are provided where possible. Queried using \textit{SIMBAD} \citep{Wenger_2000} and \textit{Aladin} \citep{Bonnarel_2000, Boch_2014}. \label{tab:comp}}
    \tablehead{\colhead{Star} & \colhead{2MASS ID} & \colhead{RA (J2000.0)} & \colhead{Dec (J2000.0)} & \colhead{B} & \colhead{V} & \colhead{J} & \colhead{H} & \colhead{K}
    }
    \startdata
        HATS-5 & J04285348-2128548 & 04:28:53.48 & -21:28:55.1 & 13.4\tablenotemark{a} & 12.6\tablenotemark{a} & 11.2 & 10.8 & 10.7 \\
        comp1 & J04285823-2122025 & 04:28:58.23 & -21:22:02.5 & - & - & 11.6 & 11.3 & 11.3\\
        comp2 & J04291347-2127062 & 04:29:13.47 & -21:27:06.2 & - & - & 11.7 & 11.5 & 11.4 \\
        comp4 & J04285582-2115027 & 04:28:55.82 & -21:15:02.8 & 13.6\tablenotemark{b} & 12.9\tablenotemark{b} & 11.6 & 11.2 & 11.1 \\
        comp5 & J04284171-2124570 & 04:28:41.71 & -21:24:57.0 & 13.9\tablenotemark{b} & 12.9\tablenotemark{b} & 11.1 & 10.6 & 10.5 \\
        comp8 & J04285532-2118520 & 04:28:55.32 & -21:18:52.0 & - & - & 12.2 & 11.9 & 11.9 \\
        comp9 & J04290165-2133321 & 04:29:01.65 & -21:33:32.2 & - & - & 12.4 & 12.2 & 12.2 \\
        comp13 & J04285764-2131228 & 04:28:57.65 & -21:31:22.8 & - & - & 12.9 & 12.6 & 12.6\\
    \enddata
    \tablenotetext{a}{\cite{Zhou_2014}}
    \tablenotetext{b}{\cite{Munari_2014}}
\end{deluxetable*}

The first transit observation, T1, used the ``Spectroscopic2'' setting, which provides unfiltered spectra covering the wavelength range $\sim$4000--9500\,\AA{}, in combination with the grism with 300 lines/mm and a blaze angle of 17.5 degrees (Gri-300-17.5). As this was still during the preliminary stages of ACCESS, the setup was changed in following years to take into account blocking filters to prevent higher-order light contamination. However, during this analysis we looked for second-order contamination in the T1 dataset and did not find any evidence for this to any significant degree, consistent with other early ACCESS studies \citep{Rackham_2017}. The T2 setup used the WB5600-9200 filter, which allows light at the same throughput in the range 5600--9200\,\AA{} with lower throughput tails down to 5350 \AA{} and up to 9350 \AA{}, although we only take advantage of the shorter wavelength extension. All wavelengths beyond this range are blocked. This was combined with the grism with 150 lines/mm and a blaze angle of 18.8 degrees (Gri-150-18.8). This combination caused there to be a significant separation between the first- and second-order light that was fully contained on the detector, which allowed for the use of T2.2. The final three transits, T3--T5, all used the same setup with the GG455 filter, which has a sharp cutoff at 4550 \AA{} and covers out to approximately 10000 \AA{}, and once again the Gri-300-17.5 disperser. Our exposure times varied from night to night, as they were optimized with respect to atmospheric conditions to maintain counts at or below half-saturation, which is what we have found works best for these precise observations \citep{Bixel_2019} and keeps us within the linearity regime of the CCD. We use $2 \times 2$ binning for all of the observations to decrease the read-out time. Our observations are summarized in \autoref{tab:obs}. 

\begin{deluxetable*}{l||c c c c c c c c}
    \centering
    \tablecaption{A summary of the observations of HATS-5 that were carried out as part of this study. (Number of comparison stars - Comp, Number of science frames - SF).\label{tab:obs}}
    \tablehead{\colhead{Observing} & \colhead{Transit} & \colhead{Exposure} & \colhead{Comp} & \colhead{SF} & \colhead{Filter} & \colhead{Disperser} & \colhead{Wavelength} & \colhead{Dispersion} \vspace{-3pt} \\
    \colhead{Date} & \colhead{Designation} & \colhead{Time (s)} & \colhead{} & \colhead{} & \colhead{} & \colhead{} & \colhead{Coverage (\AA{})} &\colhead{(\AA{}/Pixel)}
    }
    \startdata
         29 Oct 2014 & T1 & 30/45/60 & 7 & 240 & Spectroscopic2 & Gri-300-17.5 & 4000--9500 & 1.3 \\
         17 Dec 2015 & T2 & 33/30 & 7 & 381 & WB5600-9200 & Gri-150-18.8 & 5300--9350 & 2.6 \\
         17 Dec 2015 & T2.2 & 33/30 & 6 & 381 & WB5600-9200 & Gri-150-18.8 & 5300--7200 & 1.3 \\
         30 Nov 2017 & T3 & 60/50 & 5 & 200 & GG455 & Gri-300-17.5 & 4550--10000 & 1.3 \\
         20 Dec 2017 & T4 & 90 & 5 & 175 & GG455 & Gri-300-17.5 & 4550--10000 & 1.3 \\
         27 Nov 2018 & T5 & 90 & 4 & 145 & GG455 & Gri-300-17.5 & 4550--10000 & 1.3 \\
    \enddata
   
    \label{observations}
\end{deluxetable*}

\subsection{Photometric Monitoring} \label{sec:monitoring}
Stellar inhomogeneities like star spots and faculae cause an observable impact on transmission spectroscopy, even if these active regions are not directly transited \citep[e.g.,][]{Rackham_2018, Rackham:2019}. HATS-5 is a typical G-type star, and \cite{Zhou_2014} note a lack of chromospheric activity and a slow rotation rate, which points to a star with little predicted stellar activity. Nevertheless, to check for the likelihood of stellar activity impacting our results, we inspected photometric data from the All-Sky Automated Survey for Supernovae \citep[ASAS-SN,][]{Shappee_2014, Kochanek_2017}. ASAS-SN photometrically monitors the entire sky in the $V$-band with 24 telescopes worldwide. There are 320 observations of HATS-5 taken with cameras at the Cerro Tololo International Observatory in Chile and the Haleakala Observatory in Hawaii from Oct 31 2013 to Nov 27 2018, which covers our entire observation period, with an average photometric uncertainty of 18,500 ppm per datapoint. 

We detrended this data using Gaussian Processes \citep[GP, implemented with \textit{george},][]{hodlr}, and we find that the resulting data is flat to within uncertainties. Nevertheless, we looked for periodicity using a generalized Lomb-Scargle Periodogram, which excels at detecting periodic signals in unevenly sampled data \citep{Lomb_1976,Scargle_1982, vanderplas_2018}. We find two potentially significant period signals in our periodogram that predict a $\sim 5000$ ppm amplitude of variability, but with significant error bars (at around the same 5000 ppm level when phase folded and binned to 30 minutes) and corresponding unconvincing phase-folded results. Thus, we additionally looked for potential stellar activity signals in data from the Transiting Exoplanet Survey Satellite \citep[TESS, e.g.,][]{Ricker2014}, which observed HATS-5 in Sectors 5 and 32, both with 2 minute cadence. We are unable to reproduce any of the signals from the ASAS-SN data in the TESS data, and while we find two more potentially significant ($4 \sigma$) peaks in the TESS periodogram, neither of them look substantially periodic when phase-folded either. Any periodicity down to around $\sim$500 ppm should be easily identifiable in the TESS data, and so we conclude that this star is quiet at the precision of the available measurements, and that the signals we were resolving were simply due to unmodeled systematics. We also note that the star was quiet during previous RV observations \citep[rms scatter less than 5 m/s for both RV instruments,][]{Zhou_2014}, which fits with this view of a quiet host star. Nonetheless, we still model for stellar contamination as mimicking features in our transmission spectrum (see \autoref{sec:models}). Additionally, we acknowledge that \cite{Evans_2016} found a stellar companion to HATS-5, but that this companion is at a separation of $17.4 \arcsec \pm 0.06\arcsec$ from HATS-5, meaning that it would be outside even our widest slit (slit width is given in diameter, meaning that a companion would need to be within $10 \arcsec$ to be in the slit for our widest slit configuration, and a visual inspection along the slit length in an image taken before the disperser was put in place did not show any additional sources). 

\section{Data Reduction} \label{sec:reduction}

\subsection{Orbital and physical parameter updating} \label{param_updates}

An important component of performing precise transit spectroscopy involves obtaining precise and accurate orbital and 
physical parameters of the system under analysis. In particular, the orbital inclination $i$ and the semi-major axis 
to stellar radius ratio $a/R_*$ --- which in turn define the impact parameter, $b = (a/R_*) \cos i$ --- are of particular importance, as they have been shown to give rise to spurious features in the transit spectrum if not properly constrained \citep[see, e.g.,][]{alexoudi2020}. Having this picture in mind, 
we decided to update the parameters of the system using the data presented in the discovery paper by \cite{Zhou_2014}, combined with 
additional data recently obtained by the \textit{TESS} mission in Sectors 5 and 32 and precise Gaia EDR3 \citep[][]{gaia_2016, gaia_2021} parallaxes. 

Using the methods outlined in \cite{brahm1, brahm2}, we used all the available photometry for HATS-5, along with 
its spectra and Gaia parallaxes in order to derive precise stellar properties. In particular, we obtain a 
precise estimate on the effective temperature of $T_\mathrm{eff} = 5315 \pm 80$ K, a metallicity of $\mathrm{[Fe/H]} = 0.04 \pm 0.05$ 
and a surface gravity of $\log g = 4.530^{+0.022}_{-0.022}$ --- all consistent with, but more precise than, those obtained by \cite{Zhou_2014}. 
The absolute stellar mass and radius we derive from those constraints are $M_{*} = 0.868^{+0.031}_{-0.028} M_{\odot}$ and 
$R_{*} = 0.838 \pm 0.009 R_{\odot}$, which give rise to a precise stellar density of $\rho_* = 2080^{+129}_{-122}$ 
kg/m$^3$. 

Given the stellar parameters outlined above, we then perform a full fit to the entire photometric and radial-velocity 
dataset for HATS-5b. In particular, we fit the discovery HAT-South light curve and the 8 GROND light curves (two nights, 4 broadband 
light curves in the $g'$, $r'$, $i'$ and $z'$ filters obtained on each night) presented in \cite{Zhou_2014}, the full TESS 
dataset comprising two sectors worth of data, and the Subaru High Dispersion Spectrograph (HDS) and Magellan Baade Planet Finder Spectrograph (PFS) radial velocities (also from \cite{Zhou_2014}). 

Our fit is performed with \texttt{juliet} \citep{juliet}. For all the photometric datasets except for the discovery 
HAT-South light curve, we use a Gaussian Process (GP) with an multiplication between an exponential kernel and a Mat\'ern kernel to model instrumental systematics, which we detrend with respect to time. 
In particular, for the GROND data, we use wide priors on the hyperparameters of this kernel, defining a log-uniform 
prior on the timescales from 1 minute to 100 days for all bandpasses (which is a fairly wide prior given the 1.4 minute cadence of the GROND data), and a log-uniform prior on the amplitude of this GP from 1 to 1000 ppm. We also add jitter terms to capture any residual white-noise component with a log-uniform prior between 1 to 30000 ppm. We note that this is a very large upper prior limit, but we intentionally leave the bounds this wide due to the uncertainties of ground-based data. In addition, the amplitude of the GP does not necessarily correspond back to physical units.  For the TESS data, we first constrain the GP hyperparameters using the 
out-of-transit data, and use the posteriors on those fits to perform the joint fit, which only includes the in-transit 
data. We do not include a systematics model for the HAT-South light curve as these are published with a detrending 
already applied on the data. 

As for the radial-velocities, we assume these are well explained by a Keplerian orbit. 
We use a prior on the stellar density obtained in our spectroscopic analysis described above in our fits. However, following \cite{sandford} we multiply the stellar density uncertainty obtained there by two 
in our prior, in order to account for empirically determined underestimations of the uncertainty in that work. We 
tried both circular and eccentric fits, but found via the log-evidences that both models are indistinguishable by 
the data --- we find the eccentricity of the system to be $e < 0.05$ with a probability of 99\% given 
the data. 

The priors and posteriors for the planetary and stellar parameters constrained by this joint fit are presented in \autoref{tab:plprops}. The full set of priors and posteriors for all parameters can be found in a dedicated GitHub repository\footnote{\url{https://github.com/nespinoza/HATS-5-fullfit/}}. 

\begin{deluxetable*}{lcc}[t]
\tablecaption{Prior and posterior parameters of the global fit performed to HATS-5b. 
For the priors, $N(\mu,\sigma^2)$ stands for a normal distribution with mean $\mu$ 
and variance $\sigma^2$ and $U(a,b)$ stands for a uniform distribution between $a$ and $b$, respectively.\label{tab:plprops}}
\tablecolumns{3}
\tablewidth{0pt}
\tablehead{
\colhead{Parameter} &
\colhead{Prior} &
\colhead{Posterior}
}
\startdata
\multicolumn{3}{l}{\textit{Stellar \& planetary parameters}} \\
\ \ \ \ $P_1$ [d] & $N(4.7,0.1^2)$  & $4.76339146^{+0.00000046}_{-0.00000043}$ \\
\ \ \ \ $t_{0,1}$ (BJD) &  $N(2459198.5,0.1^2)$ &  $2459198.51299^{+0.00026}_{-0.00025}$ \\
\ \ \ \ $R_{p,1}/R_\star$  &  $U(0.0,1.0)$ &  $0.10620^{+0.00054}_{-0.00051}$ \\
\ \ \ \ $b_1=(a/R_\star)\cos(i)$  &  $U(0.0,1.0)$ &  $0.143^{+0.061}_{-0.063}$ \\
\ \ \ \ $K_1$ [m\,s$^{-1}$] &  $U(0,100)$ &  $30.37^{+1.09}_{-1.11}$ \\
\ \ \ \ $e_1$  &  fixed &  0 ($< 0.05$ at 99\% credibility) \\
\ \ \ \ $\omega_1$  &  fixed &  90 \\
\ \ \ \ $\rho_\star$ [kg\,m$^{-3}$] &  $N(2080,260^2)$ &  $1997^{+44}_{-58}$ \\
\multicolumn{3}{l}{\textit{TESS limb-darkening coefficients}} \\
\ \ \ \ $q_{1,TESS}^a$ &  $U(0,1)$ &  $0.47^{+0.12}_{-0.11}$ \\
\ \ \ \ $q_{2,TESS}^a$ &  $U(0,1)$ &  $0.24^{+0.12}_{-0.10}$ \\
\multicolumn{3}{l}{\textit{Ground-based photometry limb-darkening coefficients}} \\
\ \ \ \ $q_{1,HS}$ &  $U(0,1)$ &  $0.14^{+0.13}_{-0.09}$ \\
\ \ \ \ $q_{2,HS}$ &  $U(0,1)$ &  $0.25^{+0.23}_{-0.16}$ \\
\ \ \ \ $q_{1,GROND,g'}$ &  $U(0,1)$ &  $0.79^{+0.10}_{-0.10}$ \\
\ \ \ \ $q_{2,GROND,g'}$ &  $U(0,1)$ &  $0.32^{+0.07}_{-0.06}$ \\
\ \ \ \ $q_{1,GROND,r'}$ &  $U(0,1)$ &  $0.63^{+0.11}_{-0.11}$ \\
\ \ \ \ $q_{2,GROND,r'}$ &  $U(0,1)$ &  $0.18^{+0.07}_{-0.06}$ \\
\ \ \ \ $q_{1,GROND,i'}$ &  $U(0,1)$ &  $0.53^{+0.11}_{-0.10}$ \\
\ \ \ \ $q_{2,GROND,i'}$ &  $U(0,1)$ &  $0.20^{+0.08}_{-0.06}$ \\
\ \ \ \ $q_{1,GROND,z'}$ &  $U(0,1)$ &  $0.36^{+0.08}_{-0.07}$ \\
\ \ \ \ $q_{2,GROND,z'}$ &  $U(0,1)$ &  $0.32^{+0.10}_{-0.09}$ \\
\enddata
\tablenotetext{a}{These parameterize the quadratic limb-darkening law using the transformations in 
\cite{KippingLDs}.}
\end{deluxetable*}

\begin{figure*}[tbp]
    \centering
    \includegraphics[width = \textwidth]{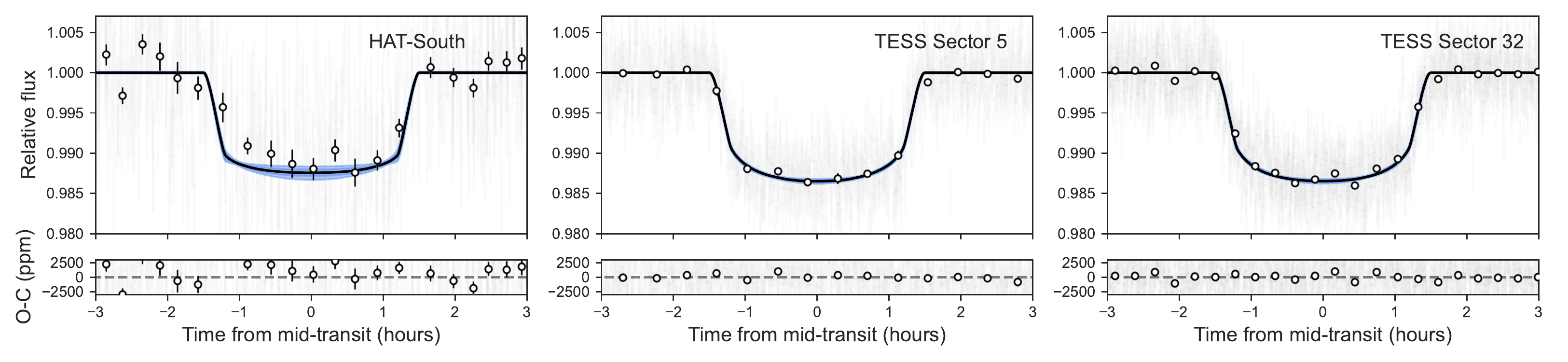}
    \includegraphics[width = \textwidth]{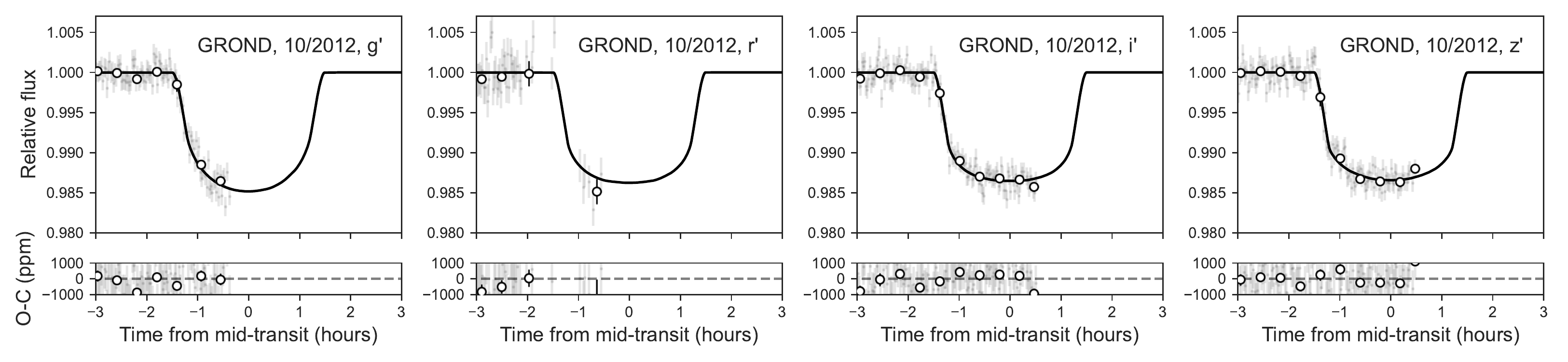}
    \includegraphics[width = \textwidth]{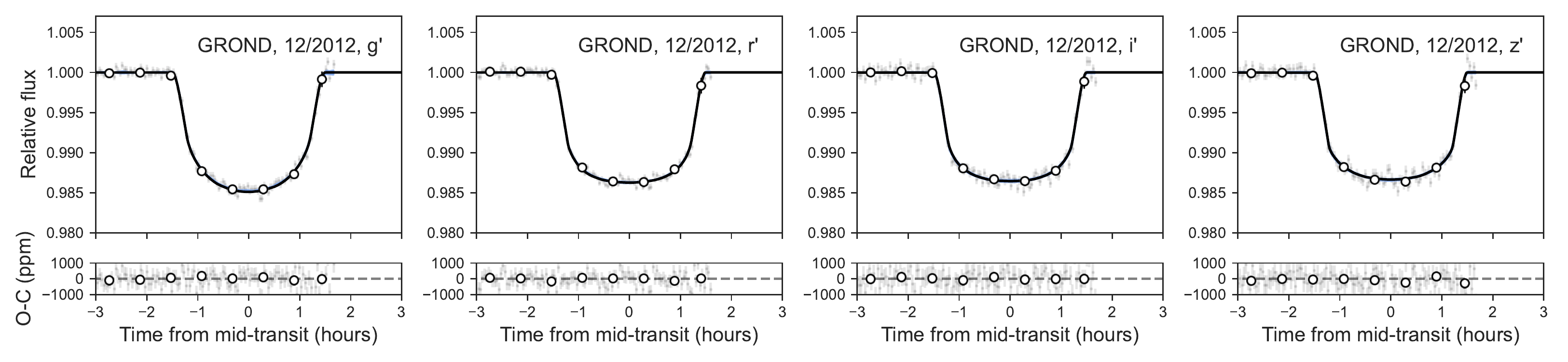}
    \includegraphics[width = \textwidth]{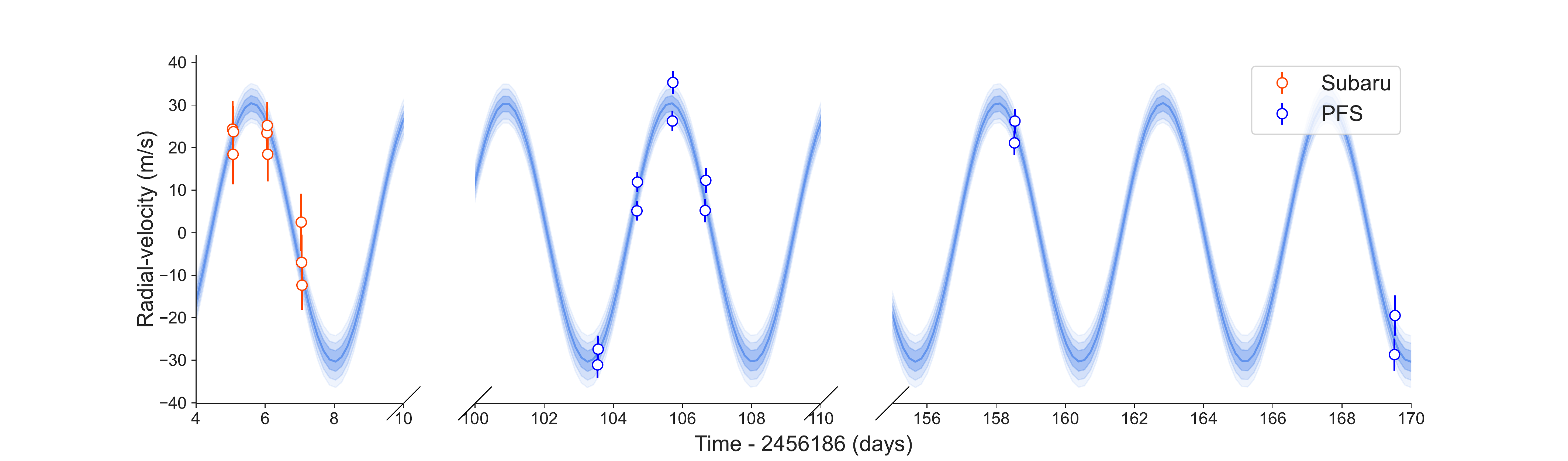}
    \caption{Results from our joint fit to all available HATS-5b photometry from HAT-South, TESS (top panels) and GROND (middle panels), along with radial-velocities from Subaru 
    and PFS (bottom panel). Blue bands denote the 68\% credibility bands of our 
    photometric models and the 68, 95 and 99\% credibility bands for 
    the radial-velocity. Note that for the photometric data the open circles correspond to binned data, shown for visualization purposes, but that all fitting was carried out for the unbinned data.}
    \label{fig:joint-fit}
\end{figure*}

This fit (\autoref{fig:joint-fit}), benefiting from the inclusion of the TESS data and our stellar density prior, significantly improves the precision of the system parameters (\autoref{tab:plprops}). First, we see an improvement of a factor of 20 on the uncertainty in the planetary period of this exoplanet, reaching sub-second accuracy (with an error of 0.04 seconds). We have been careful in 
checking that this precision is indeed warranted in principle as 
all the HAT-South data we have used are in BJD-TDB (Barycentric Julian Date-Barycentric Dynamical Time), as well as 
the newly included TESS data, which should pose a precision 
floor of $10^{-6}$ seconds from the time-stamps alone \citep{Eastman2010}. In addition, the transit parameters that define the physical properties of the system are also greatly improved. We reach a planet-to-star radius ratio precision of 0.5\%, implying a transit depth constraint based on this joint dataset of $11279\pm 115$ ppm. Interestingly, our light curve fits provide a constraint on the stellar density which is even better than the one we use as a 
prior by a factor of 4, highlighting the quality of the high-precision light curves used in our joint fit. This constraint, combined with the precisely constrained period, give rise 
to a value of $a/R_* = 13.381^{+0.098}_{-0.132}$ and $i = 89.39 \pm 0.27$ deg. Both are fully consistent with the values presented in \cite{Zhou_2014}, but are more precise by a factor of 2.5 and 1.1, 
respectively.

\subsubsection{Stellar metallicity analysis} \label{sec:stellar_met}
Exoplanet atmospheric abundances are only relevant with respect to the conditions in which they formed, and this needs to be taken into account when making any inferences from their atmospheres  \citep[][]{Reggiani_2022, molliere_2022}. While we cannot know the original protoplanetary disk composition, its chemistry is imprinted in the stellar atmosphere. Analyzing the previously mentioned high-resolution PFS data, as well as additional publicly available Fibre-fed Optical Echelle Spectrograph (FEROS) data from the MPG/ESO 2.2m telescope at La Silla Observatory, we obtain the host star abundances for HATS-5. Overall, its chemical pattern is consistent with solar, although problems with the oxygen lines covered by the data only allowed us to find abundance limits. However, given the rest of the elements' typical solar behavior, we assume solar abundance for any further conclusions from the planet's atmosphere. Further stellar abundance analysis of this host star would make these conclusions more concrete though.

\subsection{Magellan/IMACS data reduction}
We reduced the raw Magellan/IMACS data using the data reduction pipeline consistently used in ACCESS publications \citep{Jord_n_2013, Rackham_2017, Espinoza_2019, Bixel_2019, Weaver_2020, McGruder_2020, Kirk_2021, Weaver_2021, mcgruder_2022}, described in detail in \cite{espinoza_thesis_2017}, and which will be briefly summarized here. This pipeline first applies bias calibration and identifies the slit positions for each of the targets in the detector. Next, the pipeline usually corrects for bad pixels and cosmic rays, but we have omitted that process here as we observed it was giving sub-optimal results in our 
particular dataset. We thus decided to handle these outliers via optimal extraction (see \autoref{opt} below), and skipped this step in the reduction of our 
datasets. After this step, the spectra for the target and comparison stars 
are extracted for each exposure, and then the pipeline applies a wavelength calibration with the use of HeNeAr lamps that were taken on each night of observations with a narrower slit ($0.5''$) than what is used for our science frames in order to obtain higher spectral resolution. Each line profile is fit with a Lorentzian and then matched with identified lamp spectral features. Then, the profile positions are fit with a polynomial to determine the wavelength solution. This is done for the first science frame of the night, after which each following science frame is cross-correlated with the first frame's identified template to determine and correct for any wavelength offsets. It should be noted that we do not apply a flat-field correction to the data --- as with past ACCESS studies, we found that it introduced additional noise into the data rather than making a marked improvement \citep[e.g.,][]{Bixel_2019}. The result is a cleaned spectrum on a known wavelength grid for the target star and each of the comparison stars. 

\subsubsection{Optimal Extraction}\label{opt}

As described above, for our particular dataset the bad pixel and cosmic 
ray identification procedures of the ACCESS pipeline did not give optimal results, 
being unable to efficiently detect cosmic rays that significantly impacted 
the final transit light curve precisions. 
We thus decided to use optimal extraction \citep{Marsh_1989} instead to 
perform our spectral extraction, which automatically handles outliers via 
its profile fitting and weighting nature. As described in \cite{Jord_n_2013}, for most 
high signal-to-noise ratio datasets, simple and optimal extraction should 
give rise to similar results, but we found a dramatic improvement 
in our particular HATS-5b dataset. We see an improvement of between 40 to over 100 ppm in the precision of our white light curves using optimal extraction instead of the simple extraction. This might be due to the fact that 
HATS-5b has a significantly longer transit duration than any of the 
previous exoplanets analyzed by the ACCESS survey which, combined with the 
brightness of the target, give rise to better achievable precisions in the 
transit spectrum. Thus, small effects like a handful of undetected outliers 
had a more significant impact on the final results than on previous datasets. For a detailed explanation of our implementation of optimal extraction, see \citet{Brahm_2017}.

\subsubsection{Wavelength Bins}
As with previous ACCESS datasets, some of the spectra of our targets fell on two different chips, which creates a wavelength gap in the spectra.  For HATS-5b, these gaps are approximately 5990--6120\,\AA{}, no gap, 9230--9330\,\AA{}, 9230--9330\,\AA{} and 9280--9380\,\AA{} for Transits 1--5, respectively. All but the gap for Transit 1 fell redder than the wavelength coverage for Transits 1 and 2, so they would have been ignored regardless. However, we are careful to make our wavelength bins for the transmission spectra so that they do not cover the 5990--6120 \AA{} region. To determine the wavelength bins themselves, we use ExoCTK's Forward Modeling tool \citep{ExoCTK} with the Fortney grid \citep[][]{Fortney_2010} to predict the expected atmosphere of HATS-5b.\footnote{\url{https://exoctk.stsci.edu/fortney}} Using this prediction, we position bins around the most likely features---Na\,\textsc{i} and K\,\textsc{i}---and space the rest evenly around them in bins of width 150\,\AA{}. These bins are listed in \autoref{tab:bins}. The usable second-order spectrum covers 5300--7200\,\AA{}, so the bluest 11 bins have contributions from all six transits. The spectrum actually extends all the way to bin 16, but the uncertainties increase rapidly after bin 11 and so we exclude those. The rest are comprised of the other five transits, besides bin 12, which had a complex, moving, bad-pixel contamination inside of an absorption feature in T2 that we found to be uncorrectable, and thus only has 4 transits contributing to the final value. 

\begin{deluxetable*}{c c c c|| c c c c c c||c}
    \centering
    \tabletypesize{\footnotesize}
    \caption{Wavelength bins, with corresponding transit depths in ppm. Note that bin 5 only spans 140 \AA{} rather than 150 \AA{} like the rest of the bins due to the spacing requirements around the large predicted features and the chip gap. Bins 1--11 have contributions from all six transits, bin 12 from T1, T3--5, and bins 13--24 from T1--2, T3--5. \label{tab:bins}}
    \tablehead{\colhead{Bin} & \colhead{Wav start} & \colhead{Wav end} & \colhead{Transits} &\colhead{T1} & \colhead{T2} & \colhead{T2.2} & \colhead{T3} & \colhead{T4} & \colhead{T5} & \colhead{Weighted Mean} 
    }
    \startdata
        1 & 5365 & 5515 & 6 & $10583^{+259}_{-248}$ & $10929^{+510}_{-502}$ & $11093^{+1311}_{-1166}$ & $14363^{+1142}_{-1059}$ & $12324^{+816}_{-803}$ & $9280^{+1047}_{-1079}$ & $11429\pm 366$\\
        2 & 5515 & 5665 & 6 & $11194^{+221}_{-213}$ & $11939^{+588}_{-531}$ & $12301^{+608}_{-612}$ & $11292^{+914}_{-1130}$ & $10312^{+770}_{-776}$ & $10121^{+1130}_{-928}$ & $11190\pm307$\\
        3 & 5665 & 5815 & 6 & $11401^{+421}_{-477}$ & $11764^{+241}_{-232}$ & $10891^{+412}_{-366}$ & $12718^{+1733}_{-1449}$ & $11827^{+785}_{-753}$ & $10935^{+1070}_{-954}$ & $11585\pm364$\\
        4 & 5815 & 5965 & 6 & $11329^{+215}_{-230}$ & $12447^{+458}_{-455}$ & $11738^{+338}_{-340}$ & $10783^{+965}_{-892}$ & $13609^{+935}_{-817}$ & $11982^{+1349}_{-1119}$ & $11986\pm311$\\
        5 & 6120 & 6260 & 6 & $11631^{+224}_{-213}$ & $12453^{+446}_{-464}$ & $11075^{+811}_{-734}$ & $12178^{+939}_{-866}$ & $10771^{+726}_{-701}$ & $10873^{+1386}_{-1173}$ & $11500\pm320$\\
        6 & 6260 & 6410 & 6 & $10543^{+219}_{-196}$ & $11979^{+286}_{-293}$ & $11884^{+561}_{-565}$ & $11659^{+955}_{-884}$ & $11702^{+731}_{-663}$ & $11230^{+1707}_{-1362}$ & $11494\pm336$\\
        7 & 6410 & 6560 & 6 & $10755^{+225}_{-246}$ & $12534^{+669}_{-535}$ & $12015^{+705}_{-740}$ & $11140^{+1392}_{-1005}$ & $11646^{+714}_{-712}$ & $11982^{+1465}_{-1279}$ & $11684\pm367$\\
        8 & 6560 & 6710 & 6 & $10609^{+216}_{-209}$ & $12618^{+597}_{-588}$ & $11236^{+546}_{-573}$ & $12662^{+975}_{-893}$ & $12042^{+644}_{-600}$ & $10765^{+1170}_{-1059}$ & $11653\pm295$\\
        9 & 6710 & 6860 & 6 & $11388^{+1100}_{-1073}$ & $12109^{+393}_{-469}$ & $11055^{+577}_{-565}$ & $10198^{+1293}_{-1389}$ & $12005^{+805}_{-760}$ & $10411^{+1414}_{-1147}$ & $11202\pm406$\\
        10 & 6860 & 7010 & 6 & $11052^{+179}_{-173}$ & $12637^{+619}_{-620}$ & $11322^{+1400}_{-1182}$ & $11786^{+1055}_{-974}$ & $11886^{+825}_{-795}$ & $10827^{+1034}_{-906}$ & $11585\pm357$\\
        11 & 7010 & 7160 & 6 & $11571^{+212}_{-222}$ & $12269^{+519}_{-605}$ & $11495^{+964}_{-969}$ & $12169^{+966}_{-888}$ & $12365^{+867}_{-847}$ & $12679^{+1613}_{-1488}$ & $12083\pm383$\\
        12 & 7160 & 7310 & 4 & $11898^{+202}_{-226}$ & - & - & $11483^{+891}_{-804}$ & $13464^{+1365}_{-1363}$ & $12336^{+1110}_{-1107}$ & $12292\pm492$\\
        13 & 7310 & 7460 & 5 & $12105^{+345}_{-426}$ & $12387^{+716}_{-789}$ & - & $11869^{+1098}_{-1087}$ & $12064^{+991}_{-946}$ & $12098^{+1351}_{-1297}$ & $12106\pm434$\\
        14 & 7460 & 7610 & 5 & $12178^{+351}_{-332}$ & $10677^{+1198}_{-1187}$ & - & $12591^{+1241}_{-1235}$ & $12497^{+974}_{-844}$ & $11579^{+1469}_{-1248}$ & $11908\pm478$\\
        15 & 7610 & 7760 & 5 & $12134^{+361}_{-384}$ & $12426^{+598}_{-609}$ & - & $12659^{+993}_{-974}$ & $11588^{+1039}_{-966}$ & $11349^{+801}_{-868}$ & $12028\pm358$\\
        16 & 7760 & 7910 & 5 & $12771^{+434}_{-482}$ & $12256^{+649}_{-696}$ & - & $12948^{+1011}_{-989}$ & $10833^{+702}_{-675}$ & $12180^{+1472}_{-1353}$ & $12203\pm407$\\
        17 & 7910 & 8060 & 5 & $11960^{+495}_{-561}$ & $13485^{+591}_{-690}$ & - & $11939^{+941}_{-918}$ & $12589^{+1014}_{-930}$ & $10614^{+1661}_{-1302}$ & $12113\pm434$\\
        18 & 8060 & 8210 & 5 & $12512^{+267}_{-272}$ & $13108^{+536}_{-519}$ & - & $11119^{+663}_{-745}$ & $11467^{+695}_{-650}$ & $11816^{+1100}_{-1039}$ & $12008\pm313$\\
        19 & 8210 & 8360 & 5 & $12476^{+277}_{-308}$ & $12602^{+463}_{-477}$ & - & $11352^{+432}_{-468}$ & $12465^{+949}_{-991}$ & $13588^{+870}_{-931}$ & $12499\pm298$\\
        20 & 8360 & 8510 & 5 & $12270^{+318}_{-368}$ & $12170^{+393}_{-416}$ & - & $11741^{+680}_{-638}$ & $11398^{+776}_{-717}$ & $11943^{+1543}_{-1571}$ & $11901\pm389$\\
        21 & 8510 & 8660 & 5 & $12096^{+358}_{-386}$ & $11828^{+192}_{-198}$ & - & $12083^{+1051}_{-995}$ & $11977^{+975}_{-947}$ & $12579^{+1239}_{-1223}$ & $12113\pm384$\\
        22 & 8660 & 8810 & 5 & $12518^{+479}_{-560}$ & $12239^{+454}_{-579}$ & - & $11684^{+690}_{-672}$ & $10830^{+1134}_{-1063}$ & $12004^{+1030}_{-1016}$ & $11858\pm362$\\
        23 & 8810 & 8960 & 5 & $11783^{+219}_{-228}$ & $12908^{+470}_{-476}$ & - & $10781^{+798}_{-723}$ & $13641^{+1409}_{-1344}$ & $10568^{+1011}_{-989}$ & $11940\pm382$\\
        24 & 8960 & 9110 & 5 & $11813^{+261}_{-251}$ & $11972^{+254}_{-241}$ & - & $13195^{+692}_{-611}$ & $11430^{+1091}_{-1015}$ & $9382^{+1938}_{-1804}$ & $11565\pm456$
    \enddata
    \label{tab:bins}
\end{deluxetable*}

\section{Light Curves} \label{sec:lcs}
\subsection{Light Curve Fitting and Detrending}\label{sec:fitting}
Following \citet{Espinoza_2019}, we model our target star white light curve in magnitude-space as a combination of multiple elements:
\begin{equation}
    m_{T}(t) = -2.51\mathrm{log}_{10}T(t)+ m_{c}(t)+ \alpha
\end{equation}
where $m_{T}(t)$ is the magnitude of the target star, $T(t)$ is a transit model, $m_{c}(t)$ is the magnitude of (a combination of) the comparison stars, and $\alpha$ is a systematics model. 

For the transit model $T(t)$, we use \textit{batman} \citep{Kreidberg_2015}, which assumes a uniform spherical planet. For our white-light light curve fits, we fit for the period ($P$), $a/R_{\star}$, $R_p/R_{\star}$, impact parameter ($b$), and time of transit center ($t_0$). The 
priors for these parameters are set as to be consistent with the posteriors 
found in \autoref{param_updates}, except for the planet-to-star radius 
ratio and the time-of-transit center, whose uncertainties are inflated to 
allow for possible wavelength-dependent depth changes as well as 
transit timing variations. We also define the eccentricity and associated argument of periastron passage ($\omega$), but these are held constant (to zero and 90 degrees, respectively). For limb-darkening, we use the uninformative 
sampling scheme proposed by \cite{KippingLDs} for a quadratic limb-darkening law; however, following \cite{Espinoza_2016}, we use a square-root law as the limb-darkening law to use in our analysis. Our white-light priors are shown in \autoref{tab:wlc}.

For $m_c(t)$, we use principal component analysis (PCA) on our comparison 
stars to extract a set of uncorrelated signals that best explain their 
variations in time. Following the approach of \citet{Jord_n_2013}, this allows us to extract an ordered set of signals that best explain their light curves (the so-called ``principal components"), which act as linear regressors in our fit. In order to account for the fact that we do not know a-priori the optimal number of principal components to use in our fit, we follow \cite{Espinoza_2019} and perform Bayesian model averaging: we 
perform fits using 1, 2, ..., $N$ components, where $N$ is the number of comparison stars used on each dataset, and then weight the posterior distributions of each of those model fits based on their Bayesian evidences to get the final, weighted posterior distribution which now accounts for this uncertainty.

Finally, the $\alpha$ term is a systematics model used to account for any effects not corrected for by the comparison stars. This is modeled with GPs, and we use 6 regressors: the variations of airmass, wavelength-shift of our target spectrum, full-width at half maximum, sky flux, trace center position, and time. We again use \textit{george} for our GP implementation, and use a multi-dimensional squared-exponential kernel as the kernel of choice for our fits, which has given excellent results on previous ACCESS analyses \citep[e.g.][]{Weaver_2020}. The GP regressors were standardized before the fitting (i.e., they were mean-subtracted and divided by their respective standard deviations).

\begin{figure*}
    \centering
    \includegraphics[width = \textwidth]{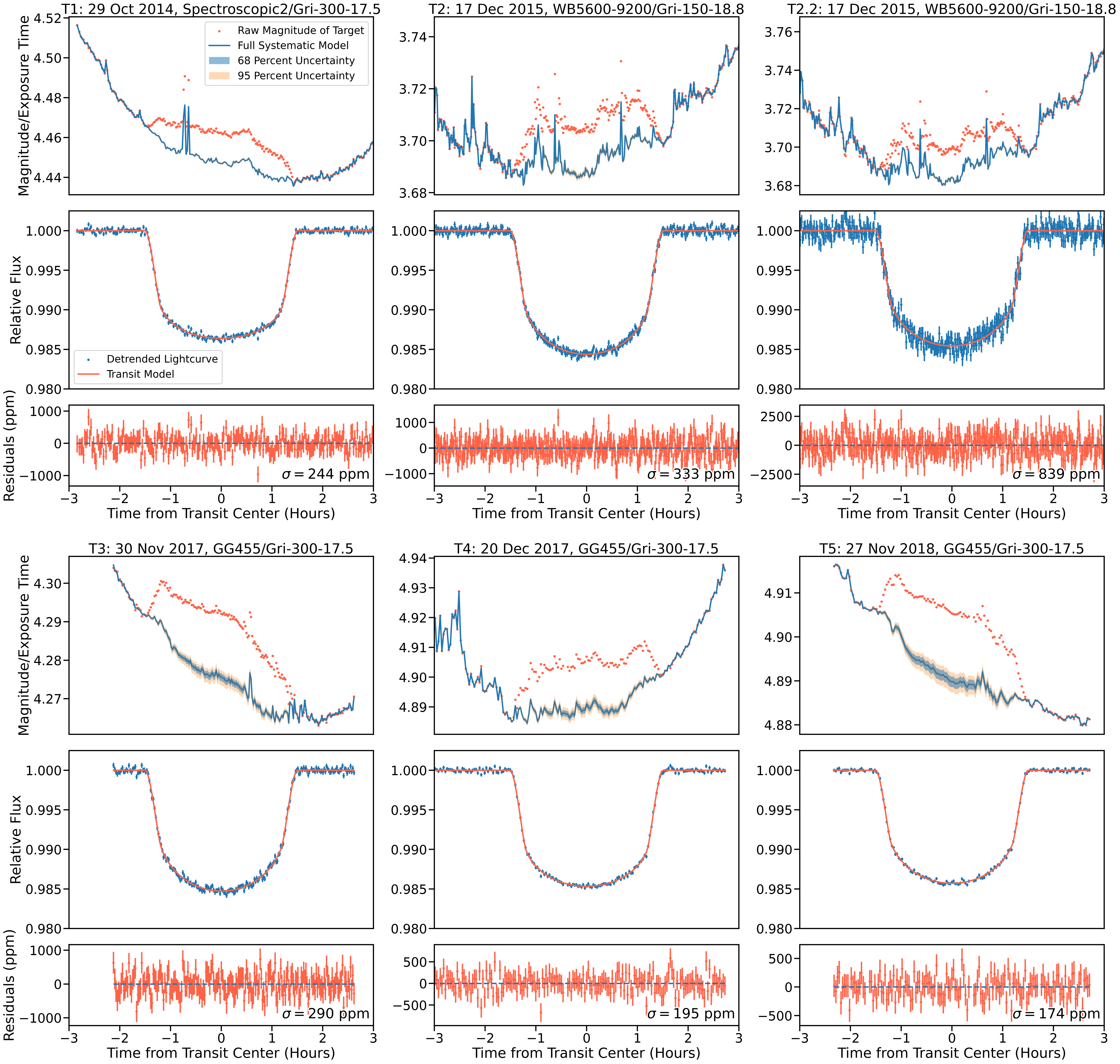}
    \caption{The results of our white light curve detrending and fitting, for each of our six transits. Top: Raw target magnitude with corresponding GP systematics model, shown with 68\% and 95\% errors. Middle: Detrended light curve in relative flux, with transit model. Bottom: Residuals with corresponding standard deviation in ppm.}
    \label{fig:wlc}
\end{figure*}

Our fits are performed via nested sampling using the \textit{pymultinest} \citep{buchner_2014} Python library, which makes use of the MultiNest algorithm \citep{Feroz_2008, Feroz_2009, Feroz_2019}. We fit each transit once, flag any flux value deviating by $\ge 5 \sigma$, and fit the light curve again not considering those 
flagged time-stamps. The results of these fits are given in \autoref{tab:wlc}, and are presented in \autoref{fig:wlc}. We obtain precisions of 174 to 333 ppm for our first-order white light curve fits, and a precision of 839 ppm for our second-order white light curve fit obtained from transit T2.2. A corner plot showing the posteriors of our white light fits are given in \autoref{app:wl_corner}. The photon noise levels for our first-order white light curves are 75 ppm, 223 ppm, 72 ppm, 75 ppm, and 73 ppm, meaning our residual standard deviation corresponds to roughly 3$\times$, 1.5$\times$, 4$\times$, 2.5$\times$, and 2.5$\times$ the photon noise for T1--T5, respectively. The photon noise level for our second-order white light curve is 342 ppm, meaning our residual standard deviation corresponds to roughly 2$\times$ the photon noise for T2.2. This is broadly consistent with the typical capabilities from the ground \cite[e.g.,][]{Rackham_2017, Espinoza_2019, Bixel_2019}.

\begin{deluxetable*}{c||c c c c c c c}
    \centering
    \caption{The priors and results for our white light curve fits. The combined final planetary fit parameters are also given, which are used for the wavelength-dependent light curve fit. \label{tab:wlc}}
    \tablehead{\colhead{} & \colhead{P} & \colhead{a/R$_\star$} & \colhead{R$_p$/R$_\star$} & \colhead{b} & \colhead{$t_0$\tablenotemark{a}}  & \colhead{q1} & \colhead{q2}}
    \startdata
         prior & $4.7633915^{+0.0000004}_{+0.0000004}$ & $13.479^{+0.059}_{-0.073}$ & $0.10628^{+0.1}_{-0.1}$\tablenotemark{b} & $0.081^{+0.056}_{-0.048}$ & $9198.51302^{+0.2}_{-0.2}$\tablenotemark{b} & 0.0-1.0 & 0.0-1.0 \\
         type & normal & normal & truncated normal\tablenotemark{c} & truncated normal\tablenotemark{c} & normal & uniform & uniform \\
         T1 & $4.7633915^{+0.0000003}_{-0.0000003}$ & $13.478^{+0.033}_{-0.037}$ & $0.1071^{+0.0006}_{-0.0006}$ &  $0.0953^{+0.0278}_{-0.0326}$ & $6959.720301^{+0.000096}_{-0.000097}$ & $0.482^{+0.062}_{-0.086}$ & $0.342^{+0.061}_{-0.089}$ \\ 
         T2 & $4.7633915^{+0.0000004}_{-0.0000004}$ & $13.447^{+0.043}_{-0.045}$ & $0.1122^{+0.0015}_{-0.0015}$ & $0.0679^{+0.0354}_{-0.0349}$ & $7383.661216^{+0.000134}_{-0.000135}$ & $0.687^{+0.181}_{-0.159}$ & $0.283^{+0.129}_{-0.157}$ \\
         T2.2 & $4.7633915^{+0.0000004}_{-0.0000004}$ & $13.487^{+0.054}_{-0.053}$ & $0.1072^{+0.0098}_{-0.0096}$ & $0.0867^{+0.0401}_{-0.0407}$ & $7383.661072^{+0.000227}_{-0.000222}$ & $0.605^{+0.183}_{-0.136}$ & $0.211^{+0.136}_{-0.124}$ \\
         T3 & $4.7633915^{+0.0000004}_{-0.0000004}$ & $13.484^{+0.045}_{-0.047}$ & $0.1094^{+0.0025}_{-0.0023}$ & $0.0842^{+0.0390}_{-0.0378}$ & $8088.642899^{+0.000173}_{-0.000177}$ & $0.763^{+0.142}_{-0.166}$ & $0.298^{+0.099}_{-0.134}$ \\
         T4 & $4.7633914^{+0.0000004}_{-0.0000004}$ & $13.465^{+0.042}_{-0.044}$ & $0.1090^{+0.0024}_{-0.0024}$ & $0.0830^{+0.0354}_{-0.0382}$ & $8107.698916^{+0.000160}_{-0.000160}$ & $0.590^{+0.220}_{-0.165}$ & $0.341^{+0.171}_{-0.196}$ \\
         T5 & $4.7633915^{+0.0000004}_{-0.0000004}$ & $13.455^{+0.040}_{-0.041}$ & $0.1081^{+0.0039}_{-0.0035}$ & $0.0719^{+0.0339}_{-0.0329}$ & $8450.660493^{+0.000128}_{-0.000142}$ & $0.565^{+0.167}_{-0.106}$ & $0.197^{+0.160}_{-0.115}$ \\
         \hline
         \hline
         Combined & $4.7633915\pm4\times10^{-7}$ & $13.469 \pm 0.0474$ & $0.1088 \pm 0.005$ & $0.0815 \pm 0.0378$ & - & - & -\\
    \enddata
    \tablenotetext{a}{As $t_0$ changes for each transit, we use the ephemerides given here to predict when the time of each transit center should in principle be, and leave the prior broad to account for any small shifts. The values given are actually $t_0-2450000$ for brevity.}
    \tablenotetext{b}{Note that the range for the priors for most variables are the errors as derived in our fit in \autoref{param_updates}, but the ranges are inflated on these parameters for the purposes of our fit. The actual errors are $\pm0.0004$ and $\pm0.00025$ for p and $t_0$, respectively.}
    \tablenotetext{c}{For our truncated normal priors, the Gaussian distribution is truncated at $\pm 1 \sigma$.}
\end{deluxetable*}

\subsection{Wavelength-dependent Light Curves}
\begin{figure*}
    \centering
    \includegraphics[width = 0.75 \textwidth]{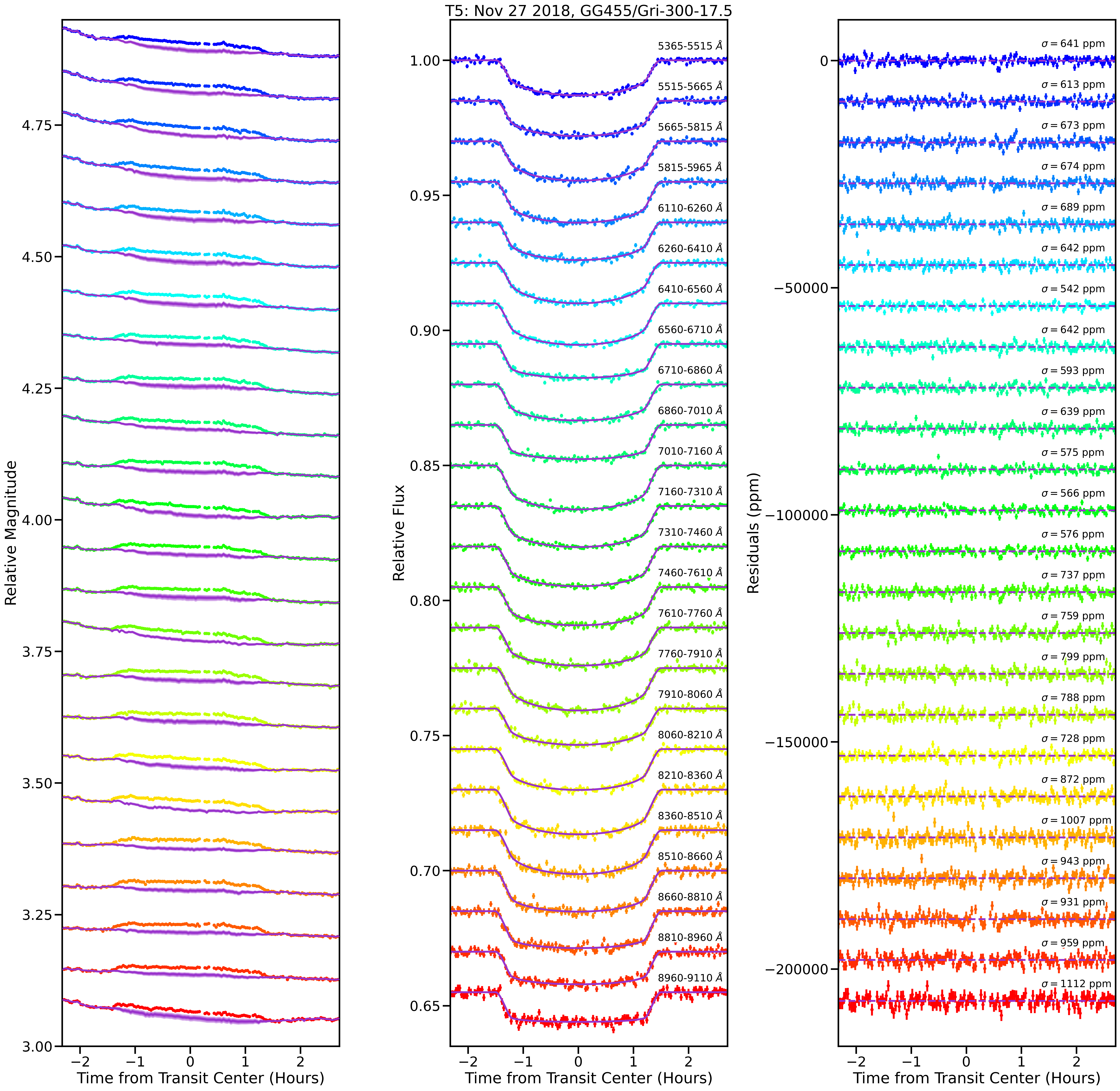}
    \caption{An example result (T5) of our wavelength-dependent curve detrending and fitting. Left: Raw target magnitude with corresponding GP systematics model, shown with 68\% and 95\% errors. Middle: Detrended light curve in relative flux, with transit model. Each wavelength bin is labeled with their wavelength range as well. Right: Residuals with corresponding standard deviation in ppm. The rest of the wavelength-dependent light curves are given in \autoref{app:wlbin}.}
    \label{fig:wavelength}
\end{figure*}

To fit the wavelength-dependent light curves, we perform the same fitting process as described in \autoref{sec:fitting}, except we hold all of the system parameters constant to the combined white light curve fit value, which is given in \autoref{tab:wlc}. The exception to this is $t_0$, which varies for each transit, and so we use the white light fit value for each transit individually for the corresponding wavelength-dependent fit. We only allow the radius ratio $R_{p}/R_\star$ to vary (up to $\pm$0.1 from the fit value) along with the limb darkening and GP terms. We complete this step for each of the 24 bins for each transit. Just as for the white light curves, we do one fit and then perform a 5$\sigma$ clip before refitting to obtain our final result. An example of the results of one transit's wavelength-dependent fits is shown in \autoref{fig:wavelength}. The rest are given in \autoref{app:wlbin}, and all of the transit depths for each wavelength bin are given in \autoref{tab:bins}. We note that the uncertainties on some of the second-order (T2.2) wavelength-dependent light curves are smaller than that for the corresponding white light curve (839 ppm), and attribute this to the systematics-dominated nature of the data. 

\subsubsection{Transmission Spectrum}
\begin{figure*}
    \centering
    \includegraphics[width=\textwidth]{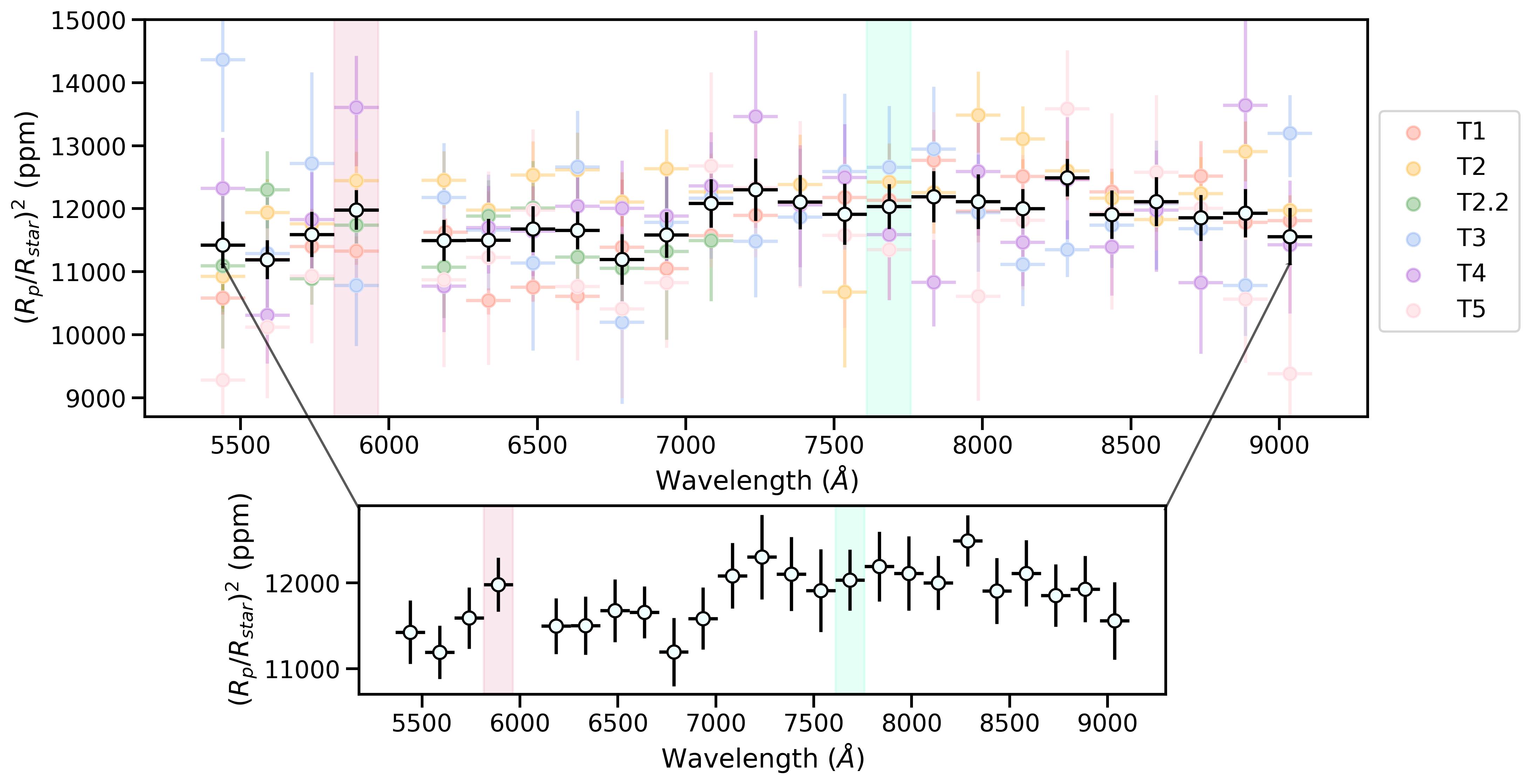}
    \caption{The final averaged transmission spectrum of HATS-5b, obtained as a weighted average of the single transit's contributions, which are also shown in the top panel. The red and green shaded regions mark the bins where Na\,\textsc{i} and K\,\textsc{i} features would be present, respectively.}
    \label{fig:spectrum}
\end{figure*}

After we complete our wavelength-dependent light curve fits, we are able to combine these fits into a final transmission spectrum through a weighted average, where for each individual spectrum we calculate 1000 samples normally distributed within the errors, and then find the mean and standard deviation of the summation of each bin's total samples for our final bin value and error. The combined spectrum of all our transits is shown in \autoref{fig:spectrum}, and the combined transit depths used are given in \autoref{tab:bins}. We note that the error bars on the second-order transmission spectrum are not notably larger than the first-order, which is understandable as we are still systematics-limited rather than photon-limited, and also that the second-order spectrum is overall consistent with our other spectra. If we take this spectra out of our final combined spectrum, the overall shape of the weighted average spectrum remains the same. However, the addition of the T2.2 spectrum decreases our error on the bluest 11 points by as much as 18 percent. Additionally, we note that unlike other ACCESS studies in which we needed to offset each transmission spectrum in order to align them, the spectrum of each night were aligned without the need for an offset. We did attempt to mean-subtract the data to align them, but we obtained the same combined transmission spectrum as without the offset. We note that this lack of offset is consistent with no significant stellar activity from the host star, as found in \autoref{sec:monitoring}. 

\section{Atmospheric Modeling}\label{sec:models}

Having obtained our combined transmission spectrum, we now switch to interpreting this in light of possible explanations for its wavelength dependence via the use of atmospheric retrievals.

\subsection{Atmospheric Retrievals}
To search for atomic and molecular signatures in our final transmission spectrum, we employ the atmospheric retrieval technique using the open-source library petitRADTRANS \citep{petitRADTRANS}. Because this library performs its sampling using Nested Sampling, in particular using the MultiNest \citep{Feroz_2008, Feroz_2009, Feroz_2019} algorithm via PyMultinest \citep{buchner_2014}, it calculates the marginal likelihood---also referred to as the Bayesian evidence---$Z=\mathcal{P}(\mathcal{D}|\mathcal{M}_i)$, i.e., the probability of the data $\mathcal{D}$ given model $\mathcal{M}_i$, for each atmospheric fit. Combined with the prior probability of each model, $\mathcal{P}(M_i)$, this allows us to compute the probability of each model given the data $\mathcal{P}(\mathcal{M}_i|\mathcal{D}) \propto Z \times \mathcal{P}(M_i)$, which enables a metric for performing model comparison between competing models attempting to explain the data. In this work, we assume all competing models are equally likely, and so perform comparisons between models by directly comparing their Bayesian evidences. In particular, following \cite{Trotta:2008}, we consider models with $\Delta$ ln Z $\leq 2$ to be indistinguishable (which roughly corresponds to a difference of $2.5 \sigma$). Before looking into more complex atmospheric models, we test the simple fit of a flat line to our data, which is consistent with the null hypothesis of no atmospheric signals detected.We calculate the Bayesian evidence for this ``Flat" model using the \texttt{Exoretrievals} package \citep{Espinoza_2019}.

As many of the spectral signatures that would allow for a determination of abundances through a free chemistry retrieval lie at redder wavelengths in the near-IR, we decide to first perform retrievals using a chemically consistent framework. The abundances of various atoms and molecules can be determined with respect to each other from just the metallicity, [Fe/H], and the carbon to oxygen ratio, C/O, for some given temperature-pressure profile if chemical equilibrium is assumed, which allows for a significant reduction in the number of free parameters in our fit. We implement this in petitRADTRANS using its subpackage \texttt{poor\_mans\_nonequ\_chem}, which interpolates abundances of defined species from a precalculated chemical equilibrium grid as a function of pressure, temperature, [Fe/H], and C/O. This grid was calculated with easyCHEM, described in \cite{molliere2017}. We implement this into our retrieval to obtain our atmospheric mass fractions and corresponding mean molecular weights for each retrieval call. All of the abundances assume the atmosphere is almost entirely composed of hydrogen and helium before these trace elements are taken into account---76.6\% H$_2$ and 23.4 \% He. We include H$_2$--H$_2$ pair collisionally induced absorption (CIA) for all of the fits, which has been argued to be important for retrieving accurate atmospheric abundances \citep{Welbanks_2019}.

Besides [Fe/H] and C/O, we allow $R_{pl}$ (reference radius), $\log \kappa_\mathrm{IR}$ (the ratio between the infrared and optical opacities), and the Guillot $\gamma$ parameter to vary. These last two parameters are required because we use the temperature-pressure profile described in \cite{Guillot:2010} to define the structure of our atmosphere. We fix $P_0$, the pressure that corresponds to $R_{pl}$, the surface gravity $\log g$, the equilibrium temperature $T_{eq}$, the internal temperature $T_{int}$, and the stellar radius $R_\star$. These are given in \autoref{tab:retparam}.

We first test the most basic model and include all of the elements and molecules that are included in the default abundance output: VO, CO, H$_2$O, C$_2$H$_2$, TiO, Na, K,  HCN, CH$_4$, PH$_3$, CO$_2$, NH$_3$, H$_2$S, and SiO. We refer to this fit as ``Consistent, Full, Clear" from here on. However, we quickly realized that the inclusion of Na and K seems to create large features not present in our data at the equilibrium parameters that fit the rest of the data. Thus, we leave them out for the remainder of our fits, assuming these are somehow depleted in the gas form in the terminator region of HATS-5b (potential explanations for this are discussed in further detail in \autoref{sec:interp}). With the remaining molecules, we obtain the atmospheric model we label ``Disequilibrium, No Na, K, Clear", as some amount of disequilibrium must be acting for these species to not be seen (such as condensation or rainout). While we include all of these molecules for completeness and for potential future predictions, we note that some of the molecules do not seem to be contributing to the final predicted atmospheric spectrum, as their opacities are simply flat in this region. However, this additional complexity does not affect our corresponding $\ln Z$ as the fit values are still only [Fe/H] and C/O, which we confirmed via testing the exclusion of various flat molecules and seeing no change in the $\ln Z$. 

Thus far, we have ignored the potential effect of clouds, so we additionally test a fit including $\log P_\mathrm{cloud}$ (the pressure at which an opaque cloud deck is added to the absorption opacity), referred to as ``Disequilibrium, no Na, K, Cloudy." These retrievals that fit for [Fe/H] and C/O we refer to collectively as our chemically consistent fits.

We note that in our fits thus far, it seems that the major spectral contribution is due to H$_2$O. Thus, to test the detection of H$_2$O, we  perform a series of free retrievals on our data using only the parameters that may be potentially important in our spectrum --- the alkalis (Na and K), H$_2$O, and clouds. As above, we also include CIA. As suggested in \cite{Welbanks_2019}, the inclusion of CIA has the potential to guard against biases related to determining abundances without including all the molecules in a fit. Besides the abundances, we leave $R_{pl}$ free, as before, as well as the temperature. Rather than the Guillot profile from the consistent model, the free retrievals use an isothermal temperature profile, so only one parameter is required. We fix $\log g$ and $R_\star$, just as in the chemically consistent case. These models are labeled as ``Alkalis + H$_2$O + Cloud", ``H$_2$O + Cloud", ``Cloud", ``Alkalis + H$_2$O", and ``H$_2$O."

\subsection{Stellar Contamination}

We also consider the possibility that any spectral features in our transmission spectrum could be generated by stellar contamination \citep{Rackham_2018, Rackham:2019}, i.e., by heterogeneities on the star rather than by only the planetary atmosphere. We note, however, that this model seems unlikely a-priori to explain the data given that \cite{Rackham:2019} predicts stellar contamination signatures to be relatively small in amplitude for planets orbiting Sun-like stars (typical G stellar type with approximately solar metallicity) like HATS-5. This prior is further strengthened in this case as HATS-5 seems to be a relatively inactive star at least from spectroscopic indicators \citep{Zhou_2014} and photometric 
monitoring of the star (see \autoref{sec:monitoring}), as well as a lack of activity indicators in our own analysis of high resolution observations (PFS, FEROS) (see \autoref{sec:stellar_met}). Nevertheless, we test the potential effects of stellar heterogeneities mimicking an atmospheric signal using the \texttt{Exoretrievals} package. The stellar 
contamination within the framework of that work is modelled following \cite{Rackham_2018}, and essentially determines the fraction of any stellar heterogeneities and the resulting effect on a transmission spectrum of choice. We allow for both temperatures less than and greater than the stellar photosphere, corresponding to potential spots and faculae respectively. We model it with an underlying flat line model, to test the idea of this spectrum being purely mimicked by stellar activity.  We refer to this model as ``Flat + Stellar Contamination".



\begin{deluxetable*}{l l c}[h!]
    \centering
    \caption{Each of our retrievals, given with a short description and their corresponding $\Delta \ln Z$ from our best-fit model ($\ln Z = -7.9$), H$_2$O, which is bolded. This and our best-fit chemically consistent retrieval (``Disequilibrium, No Na, K, Clear")}, given the $\ln Z$, are both able to rule out the null-hypothesis flat line case.  \label{tab:ret}
    \tablehead{\colhead{Model} &\colhead{Description} & \colhead{$\Delta$ ln Z}}
    \startdata
    \multicolumn{3}{|c|}{Chemically Consistent Retrievals} \\
    \hline
        Disequilibrium, No Na, K, Clear & CIA and disequilibrium chemistry retrieval excluding Na and K without clouds & 1.4 \\ [.1 cm]
        Flat & flat line retrieval at median $R_p/R_s$ testing lack of atmospheric features & 5.5 \\[.1 cm]
        Disequilibrium, No Na, K, Cloudy & CIA and disequilibrium chemistry retrieval excluding Na and K with clouds & 5.6 \\ [.1 cm]
        Flat + Stellar Contamination & stellar inhomogeneities with underlying flat line retrieval testing mimicked atmospheric features& 6.0 \\[.1 cm]
        Consistent, Full, Clear & chemically consistent retrieval including all default included elements and molecules & 8.9 \\[.1 cm]
        \hline
        \multicolumn{3}{|c|}{Free Abundance Retrievals} \\
        \hline
        \textbf{H$_2$O} & \textbf{CIA and H$_2$O free retrieval} & \textbf{0.0} \\ [.1 cm]
        Alkalis + H$_2$O & CIA and H$_2$O, Na, K free retrieval & 2.6 \\ [.1 cm]
        H$_2$O + Clouds & CIA and H$_2$O free retrieval with clouds & 2.9 \\ [.1 cm]
        Alkalis + H$_2$O + Clouds & CIA and H$_2$O, Na, K free retrieval with clouds & 4.0 \\ [.1 cm]
        Clouds & CIA and clouds & 6.1 
    \enddata

\end{deluxetable*}

\begin{deluxetable*}{l||l c|| c}
    
\centering
    \caption{The parameters used in our best-fit atmospheric models, given with their description and prior, as well as their retrieved fit value. The fixed parameters are also included with their fixed value. \label{tab:retparam}}
    \tablehead{\colhead{Parameter} & \colhead{Description} & \colhead{Prior (uniform)} & \colhead{Fit Value}}
    \startdata
         \multicolumn{4}{|c|}{Best-Fit Chemically Consistent: Disequilibrium, No Na, K, Clear} \\
         \hline
         $R_{pl}$ & reference radius & [0.8, 1.0] $R_J$ & $0.92\pm0.01$ $R_J$ \\
         C/O & atmospheric carbon to oxygen ratio & [0.2, 1.7] & $0.45\pm 0.27$ \\
         $\mathrm{[Fe/H]}$ & atmospheric metallicity relative to solar & [-2.0, 1.0] & $-0.5 \pm 1.0$ \\
         $\log \kappa_{IR}$ & ratio between infrared and optical opacities & [-5,0] & $-2.3 \pm 1.6$ \\
         $\gamma$ & Guillot \textit{gamma} parameter & [0.0, 0.8] & $0.46 \pm 0.23$ \\
         $P_0$ & pressure corresponding to $R\_pl$ & [0.01] & - \\
         $\log g$ & planetary surface gravity & [2.85] & - \\
         $T_{eq}$ & planetary equilibrium temperature & [1025] K & - \\
         $T_{int}$ & planetary internal temperature & [200] K & - \\
         $R_\star$ & stellar radius & [0.838] $R_\odot$ & - \\
         \hline 
         \multicolumn{4}{|c|}{Best-Fit Free Abundance: H$_2$O} \\
         \hline
         $R_{pl}$ & reference radius & [0.8, 1.0] $R_J$ & $0.92\pm0.02$ $R_J$ \\
         Temperature & isothermal atmospheric temperature & [300, 1900] K & $985\pm 313$ \\
         H$_2$O & atmospheric water abundance & [-6.0, 6.0] & $-3.0 \pm 1.5$ \\
         $\log g$ & planetary surface gravity & [2.85] & - \\
         $R_\star$ & stellar radius & [0.838] $R_\odot$ & - \\
    \enddata

\end{deluxetable*}

\begin{figure*}
    \centering
    \includegraphics[width=\textwidth]{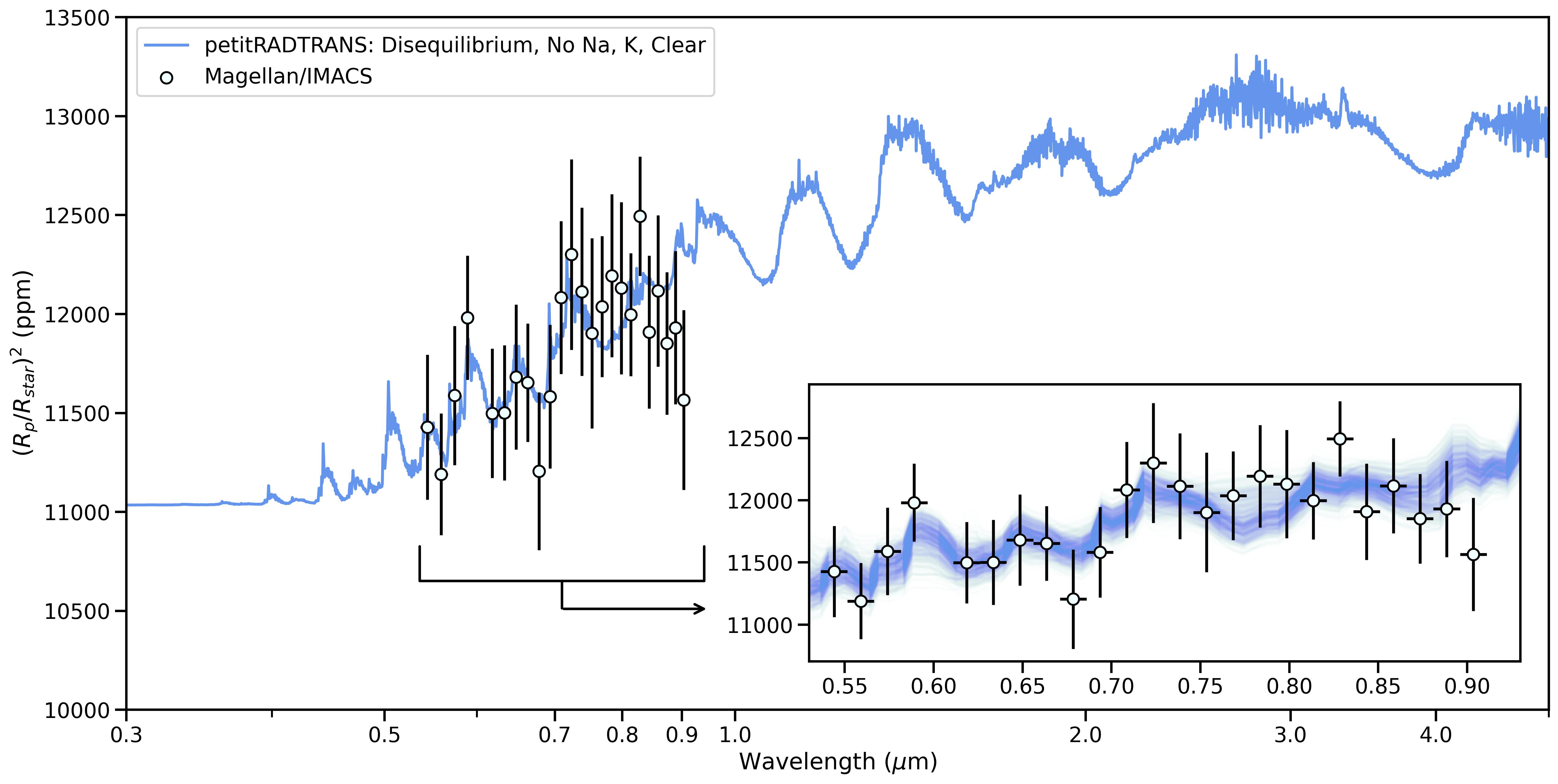}
    \caption{The best-fit retrieval for our data, the Disequilibrium, No Na, K, Clear case, plotted with the transmission spectrum. Shown in the zoomed-in box in the lower left is 1000 samples of our retrieval. The retrieval is extended down to 0.3 $\mu$m and out to 5 $\mu$m to show features in the UV and near-IR, which are observable with HST and JWST. }
    \label{fig:retrieval}
\end{figure*}

\subsection{Retrieval Interpretation}\label{sec:interp}
The results of all our retrievals are given in \autoref{tab:ret}. Out of all the models tested, our best-fit (i.e., the one for which we attain the largest Bayesian evidence) chemically consistent model is the ``Disequilibrium, No Na, K, Clear" case, and our best free retrieval is the ``H$_2$O" case.  This best-fit chemically consistent retrieval is plotted against the data in \autoref{fig:retrieval}, and the resulting retrieved parameters for both this and the ``H$_2$O" case are given in \autoref{tab:retparam}. The ``Flat", ``Flat + Stellar Contamination", ``Consistent, Full", and ``Disequilibrium, No Na, K, Cloudy" models all have $\Delta \ln Z \gtrsim 4$ when compared against our chemically consistent best-fit model. That is, given the data, those models are more than $\exp( \ln Z)\approx 50$ times less likely than our best-fit model (assuming the models themselves are \textit{a priori} equally likely). This lends moderately strong evidence to our best-fit model over those models. Notably, the higher log-evidence of the ``Disequilibrium, No Na, K, Clear" retrieval against the ``Disequilibrium, No Na, K, Cloudy" retrieval suggests a preference for a lack of clouds at the altitudes probed in transmission for HATS-5b, which is consistent with the theory for a ``clear" window around 1000 K proposed by \citet[][]{Gao_2021}. A plot showing all of the retrievals is given in \autoref{fig:retrieval_full} in \autoref{app:ret}.

Our best-fit chemically consistent model parameters are broadly consistent with expectations. The C/O ratio is sub-solar, and less than 1 at ${\sim}3 \sigma$, and the metallicity is approximately solar, although relatively unconstrained. The temperature-pressure profile, which is given solely by the Guillot profile parameters considering that we fix the temperature parameters, is also somewhat unconstrained but consistent with expectations.  

From our best-fit free model (``H$_2$O"), which is almost identical in appearance to the chemically consistent best-fit model, we are able to determine the molecular contributions to the features in the consistent model. We derive the abundances of the fit H$_2$O to be $\log(X_{\mathrm{H_2O}}) = -0.78\pm0.56$. Thus, we constrain H$_2$O to a non-zero value as our best-fit model, even more likely than the best-fit chemically consistent model. The ``H$_2$O" is a better fit than the ``Flat" by $\Delta \ln Z = 5.5$, roughly corresponding to 3.6$\sigma$. We note that several of the free retrievals are essentially indistinguishable from each other given the evidence, but that all those with $\Delta \ln Z < 3$ include the presence of H$_2$O. The retrieved temperature from this free retrieval is quite uncertain ($985 \pm 313$ K), but consistent with the equilibrium temperature of the planet (1000 K). Corner plots of the retrieved parameters from both our best chemically consistent and free chemistry retrievals are given in \autoref{fig:corner_consistent} and \autoref{fig:corner_free} of \autoref{app:ret} respectively.

From the results of our best fits, we report a tentative detection of H$_2$O in HATS-5b. H$_2$O is already known to be common in the atmospheres of hot giant planets, and is a common detection in the near-IR \citep[see e.g.][]{Sing_2015}, but not in the optical. Notably, \cite{Stevenson_2016} found evidence for H$_2$O in the optical spectrum of HAT-P-26b with LDSS-3C at Magellan/Clay. However, we note that the spectrum is consistent enough with a flat line ($\Delta \ln Z \lesssim$4--6, depending if compared against our best-fit free or chemically consistent retrieval) that we consider this evidence for water alone to not be significant enough and thus in need of confirmation. 

The lack of Na and K atomic features in the transmission spectrum of HATS-5b is, however, a puzzle. Our disfavoring of all of the retrievals including clouds suggests that they are not masked by clouds, but instead likely depleted in the atmosphere as suggested by the indistinguishability of the ``Alkalis + H$_2$O" free abundance retrieval, but strong disfavoring of the ``Consistent, Full, Clear" retrieval which includes the alkalis at their predicted abundance from equilibrium chemistry. From our analysis of the stellar abundances (\autoref{sec:stellar_met}), the abundances of Na and K in the host star seem consistent with solar, so we can rule out an inherent depletion of these elements in the system. One possible pathway to explain it could be for these elements to be trapped in molecular form. It is predicted that Na and K condense at ${\sim}900-1000$ K at pressures of $10^{-2}-10^{-3}$ (which is where we probe with transmission spectroscopy), into Na$_2$S and KCl, at which point they are sequestered away out of the optical transmission spectrum \citep{Lodders_1999, Madhusudhan_2016, Line_2017}. However, we note that these condensation relationships are complicated and somewhat uncertain, and that there may be another explanation for the lack of these features. Further chemical condensation and atmospheric structure models could dive deeper into these possibilities to explain the lack of Na and K features HATS-5b transmission spectrum. It is also possible that the resolution level probed by our observations is not high enough to identify sharp alkali features. Higher resolution observations would be able to uncover if this is the case.

\subsubsection{Potential Follow-Up Observations}
As demonstrated in \autoref{fig:retrieval}, our chemically consistent best-fit model for the optical data predicts large water features in the near-IR, which would be observable with \emph{HST}, as has been done often in the literature. For example, \cite{Wakeford_2017} followed up on the previously described optical H$_2$O evidence in HAT-P-26 b with HST observations using STIS 750L, WFC3 G102, and WFC3 G141 (0.56--1.62 $\mu$m), and were able to obtain a precise H$_2$O abundance for the planet. While our visible water features combined with the potential near-IR features would confirm the result, and allow for a more precise absolute water abundance determination, our simulations show that the result of \emph{HST} observations would not constrain the general atmospheric parameters of the planet much more. We simulated one transit observation with WFC3/UVIS G280 (0.2--0.8 $\mu$m) and one with WFC3/IR G141 (1.1--1.7 $\mu$m) using PandExo \citep{pandexo}, based on our best-fit model result shown in \autoref{fig:retrieval}, and found that we would not constrain C/O significantly more, and would only improve our estimate of [Fe/H] by a factor of 1.5.

\begin{figure*}
    \centering
    \gridline{\fig{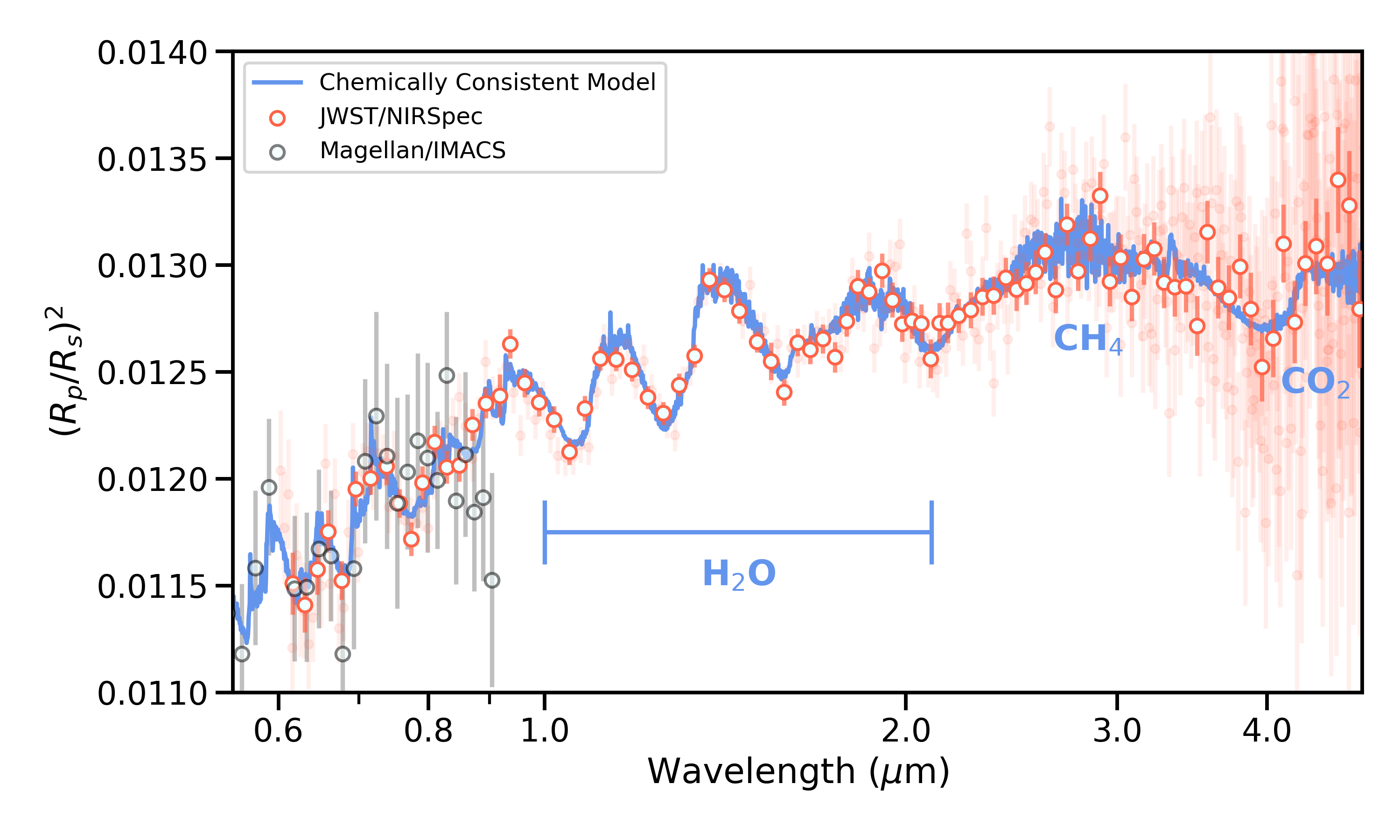}{0.6\textwidth}{}
    \fig{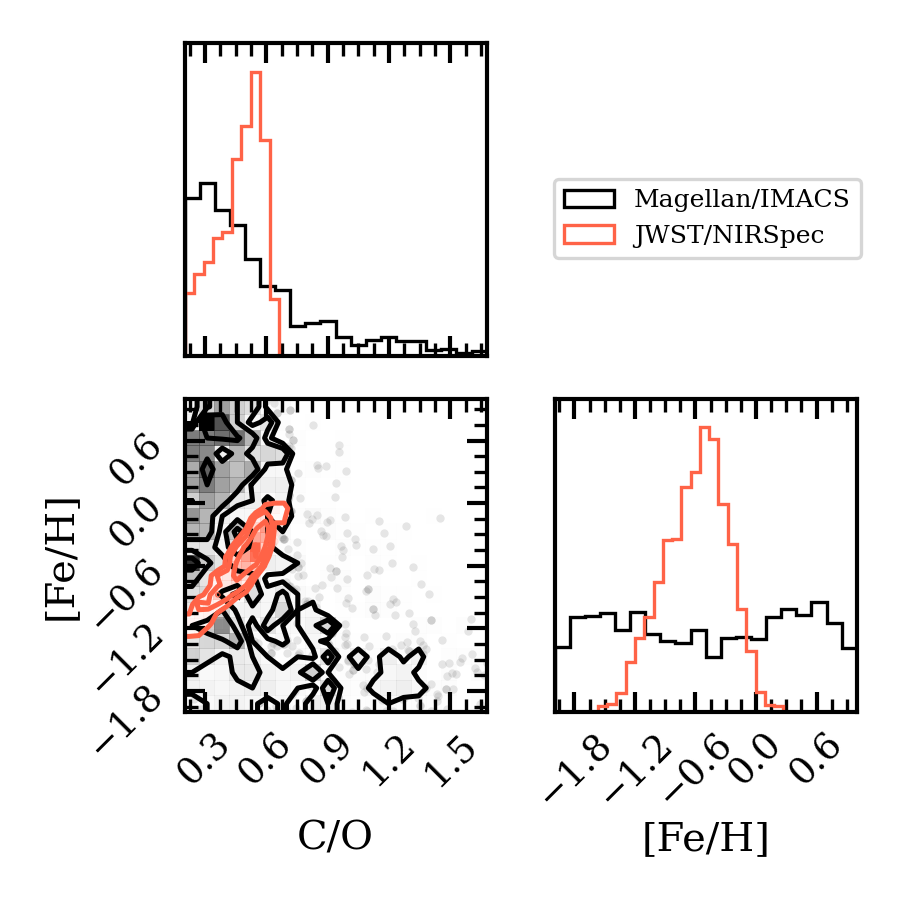}{0.35\textwidth}{}
    }
    \caption{Left: Simulated observations of HATS-5b with JWST NIRSpec PRISM based on our best-fit Disequilibrium, No Na, K, Clear model, shown with the Magellan/IMACS results from this paper. Note that the red open circles correspond to binned JWST data, shown for visualization purposes, but that all fitting was carried out for the unbinned data. We label features of interest as well. Right: The change in our posteriors between the Disequilibrium, No Na, K, Clear retrieval (the underlying model fed into the JWST simulations) and the new JWST retrieval, which illustrates the significant improvement in our determination of atmospheric parameters.}
    \label{fig:jwst}
\end{figure*}

However, observations with \emph{JWST} would further extend our potential detections beyond H$_2$O to other molecules in the near-IR, which would significantly improve our understanding of this planet. In particular, observations with NIRSpec PRISM would cover near-IR molecular features other than water, such as CO$_2$ and CH$_4$, that would better constrain atmospheric parameters like C/O, as well as being able to confirm H$_2$O better than WFC3/IR G141 could. We simulated NIRSpec PRISM observations with PandExo \citep{pandexo} and found that we could constrain these features with observations of just one transit. We show this in \autoref{fig:jwst}. We improved the precision of our C/O and [Fe/H] determinations from our chemically consistent retrievals by 2$\times$ and 3$\times$, respectively, with this additional data (to $0.49 \pm 0.12$ and $-0.58 \pm 0.33$, respectively), and increase our abundance retrieval of H$_2$O by a factor of 2 (in log space) with the corresponding free abundance retrievals.

\section{Conclusions} \label{sec:conclusions}
In this paper, we report five separate optical transits of the planet HATS-5b, obtained with the Magellan/IMACS multi-object spectrograph between 2014 and 2018. Additionally, we introduce the use of the second-order light to obtain an additional transit and apply the method to one of our observations (the 2015 transit), which was possible due to the observing setup used. With these six transits, we obtain white light curves with precisions ranging from 174 to 333 ppm for the first-order, and 839 ppm for the second-order.

The transits cover a common wavelength range of 5365--9110 \AA{}, which we bin into 150 \AA{} with an average precision per bin of 375 ppm. We perform broad atmospheric retrievals on the resulting combined transmission spectrum with Exoretrievals and petitRADTRANS, considering all cases, including a flat line (no atmospheric features detected), stellar contamination, and a variety of chemically consistent and free abundance atmospheric cases. Our highest-evidence case points to an approximately solar-metallicity atmosphere with sub-solar C/O, although the metallicity as well as the associated temperature-pressure profile remain relatively unconstrained. Using a free-retrieval, we are able to determine that most of the predicted fit features can be explained by H$_2$O, and so we report a tentative detection of this molecule. The presence of these features agree with the prediction of ``clear" atmospheric conditions around 1000 K, which is predicted to lie between the regions strongly influenced by either hydrocarbon hazes below or silicate clouds above. 

While a moderately strong detection according to our atmospheric retrievals, we still consider this detection of H$_2$O in the optical a tentative detection that needs confirmation due to the relatively small deviations from a flat line model. Our best-fit model predicts large H$_2$O features in the near-IR, making HATS-5b a strong target for atmospheric follow-up with \emph{JWST}. Not only would the observation of H$_2$O features with \emph{JWST} bolster the credibility of this optical detection, but it would also allow for a precise determination of the absolute H$_2$O abundance in its atmosphere as well as a constraint on the C/O ratio in the atmosphere from the availability of additional molecular features in the covered wavelength range. These atmospheric parameters will in turn offer clues regarding the formation and subsequent orbital evolution of HATS-5b, and contribute to the fuller picture surrounding hot Jupiters and their mysteries. 

\begin{acknowledgments}
We thank our anonymous reviewer for their thoughtful and constructive comments and suggestions. This material is based upon work supported by the National Science Foundation Graduate Research Fellowship under Grant No. DGE1746891 (N.H.A) and No. DGE1745303 (C.M. and K.O.C.), and supported by a Ford Foundation Predoctoral Fellowship (K.O.C.). B.V.R. thanks the Heising-Simons Foundation for support. The results reported herein benefited from collaborations and/or information exchange within NASA’s Nexus for Exoplanet System Science (NExSS) research coordination network sponsored by NASA’s Science Mission Directorate. This material is partly based upon work supported by the National Aeronautics and Space Administration under Agreement No. 80NSSC21K0593 for the program “Alien Earths”. H.R. acknowledges support from a Carnegie Fellowship. R.B. acknowledges support from FONDECYT project 11200751 and from ANID - Millenium Science Initiative - ICN12\_009. This paper includes data gathered with the 6.5 meter Magellan Telescope located at Las Campanas Observatory, Chile. We thank the staff at the Magellan Telescopes and Las Campanas Observatory for their ongoing input and support to make the ACCESS observations presented in this work possible.
\end{acknowledgments}

%

\vspace{5mm}
\facilities{LCO (Baade 6.5m, IMACS), TESS, ASAS-SN, GROND, Subaru, HAT-South}


\software{Aladin \citep{Bonnarel_2000, Boch_2014}, astropy \citep{2013A&A...558A..33A,2018AJ....156..123A}, batman \citep{Kreidberg_2015}, corner \citep{corner}, george \citep{hodlr}, GPTransmissionSpectra (\url{https://github.com/nespinoza/GPTransmissionSpectra}), Juliet \citep{juliet}, jupyter \citep{Kluyver2016jupyter}, matplotlib \citep{Hunter:2007}, multinest \citep{Feroz_2008, Feroz_2009, Feroz_2019}, NumPy \citep{harris2020array}, pandas \citep{reback2020pandas, mckinney-proc-scipy-2010}, pymultinest \citep{buchner_2014}, petitRADTRANS \citep{petitRADTRANS}, PandExo \citep{pandexo},  SciPy \citep{2020SciPy-NMeth}, SIMBAD \citep{Wenger_2000}
          }


\bibliography{biblio}{}
\bibliographystyle{aasjournal}

\pagebreak
\appendix

\section{White Light Curve Corner Plot}\label{app:wl_corner}
\begin{figure*}[b]
    \centering
    \includegraphics[width = \textwidth]{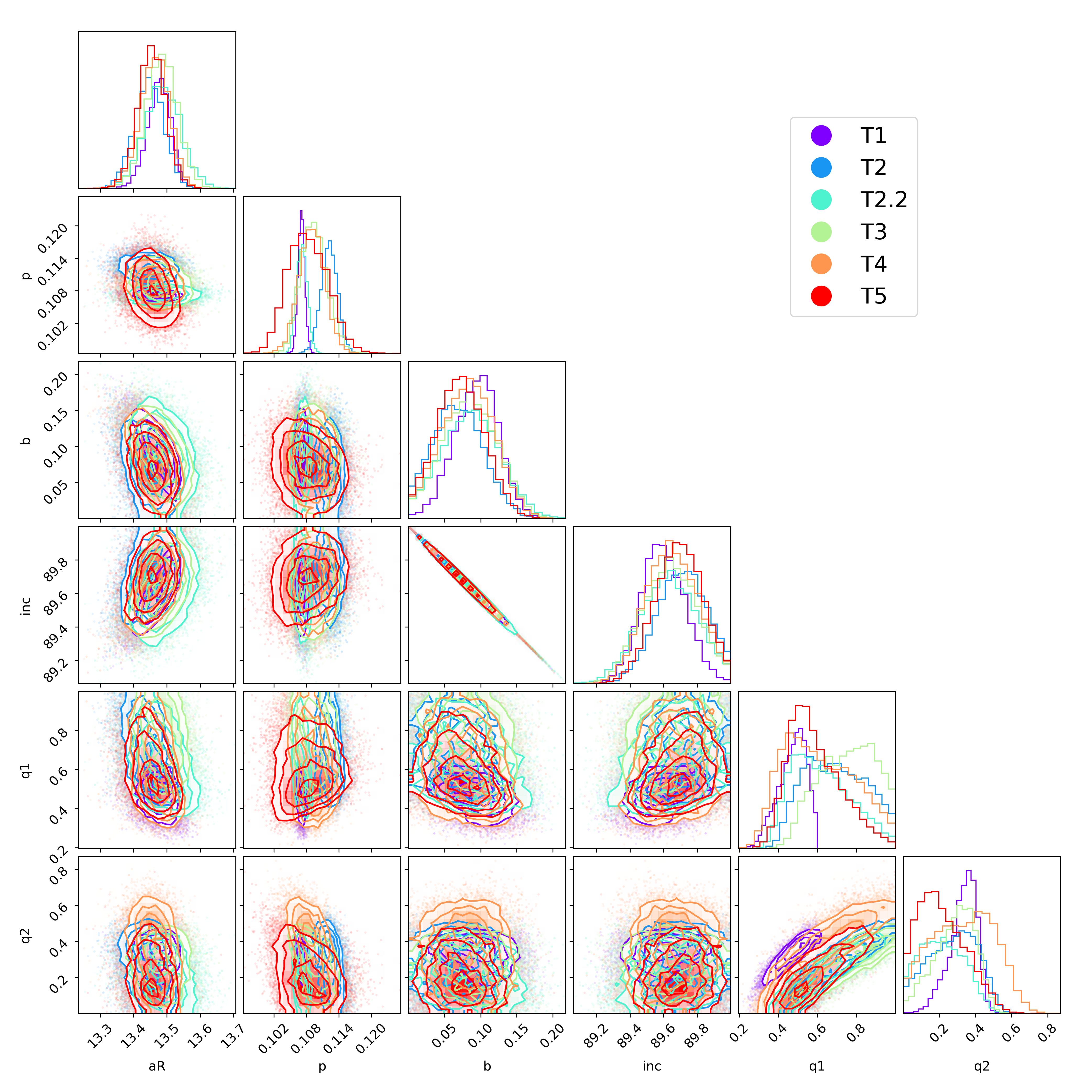}
    \caption{Corner plot showing the posterior distributions for the parameters shown in \autoref{tab:wlc} for all of the transits.}
    \label{fig:wl_corner}
\end{figure*}

\autoref{fig:wl_corner} shows a corner plot representation of our white light curve fit posteriors.

\pagebreak
\section{Wavelength Binned Light Curves}\label{app:wlbin}
\begin{figure*}[b]
    \centering
    \includegraphics[width = \textwidth]{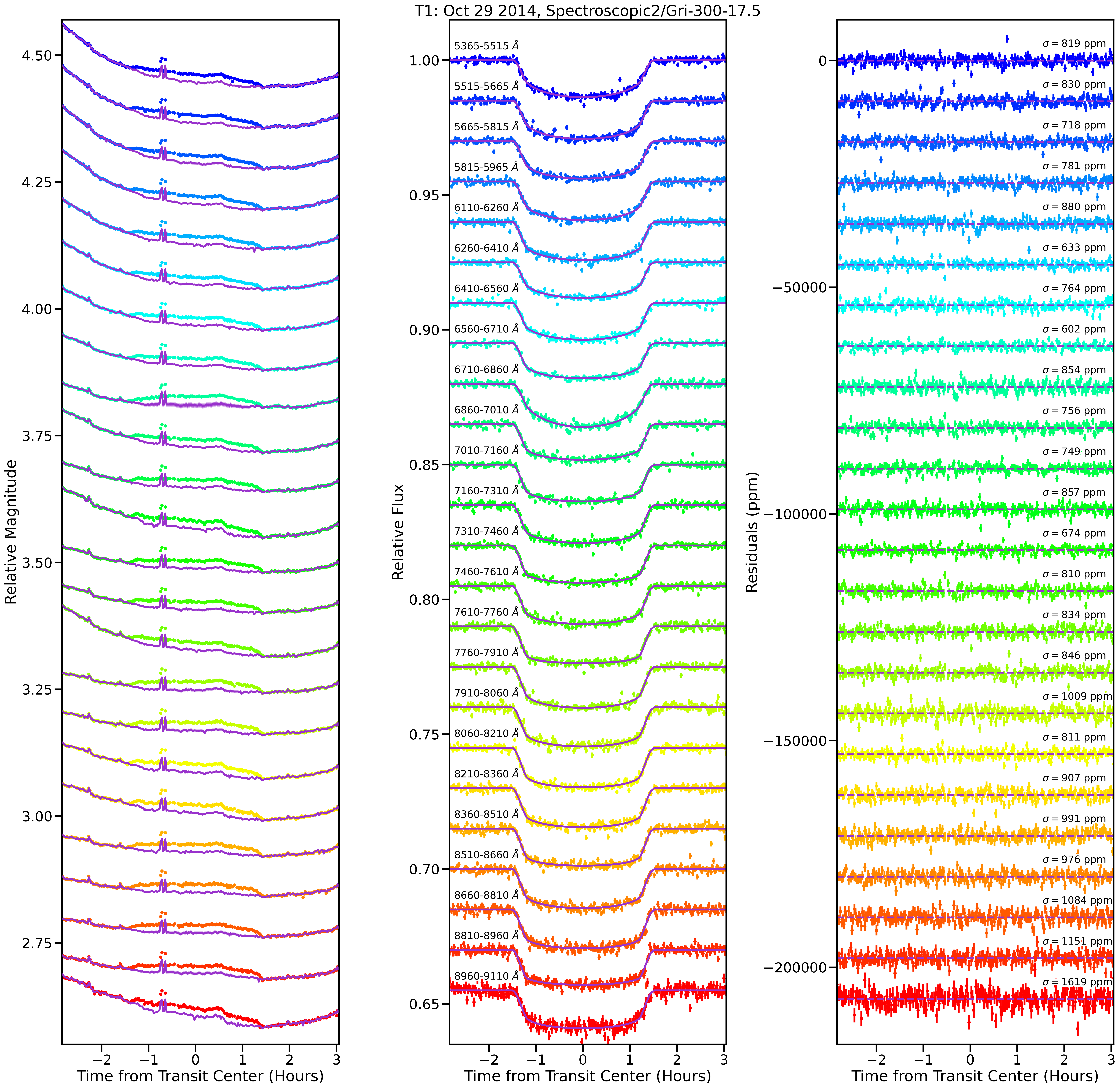}
    \caption{The same as \autoref{fig:wavelength}, but for T1.}
    \label{fig:wlT1}
\end{figure*}

\begin{figure*}[p]
    \centering
    \includegraphics[width = \textwidth]{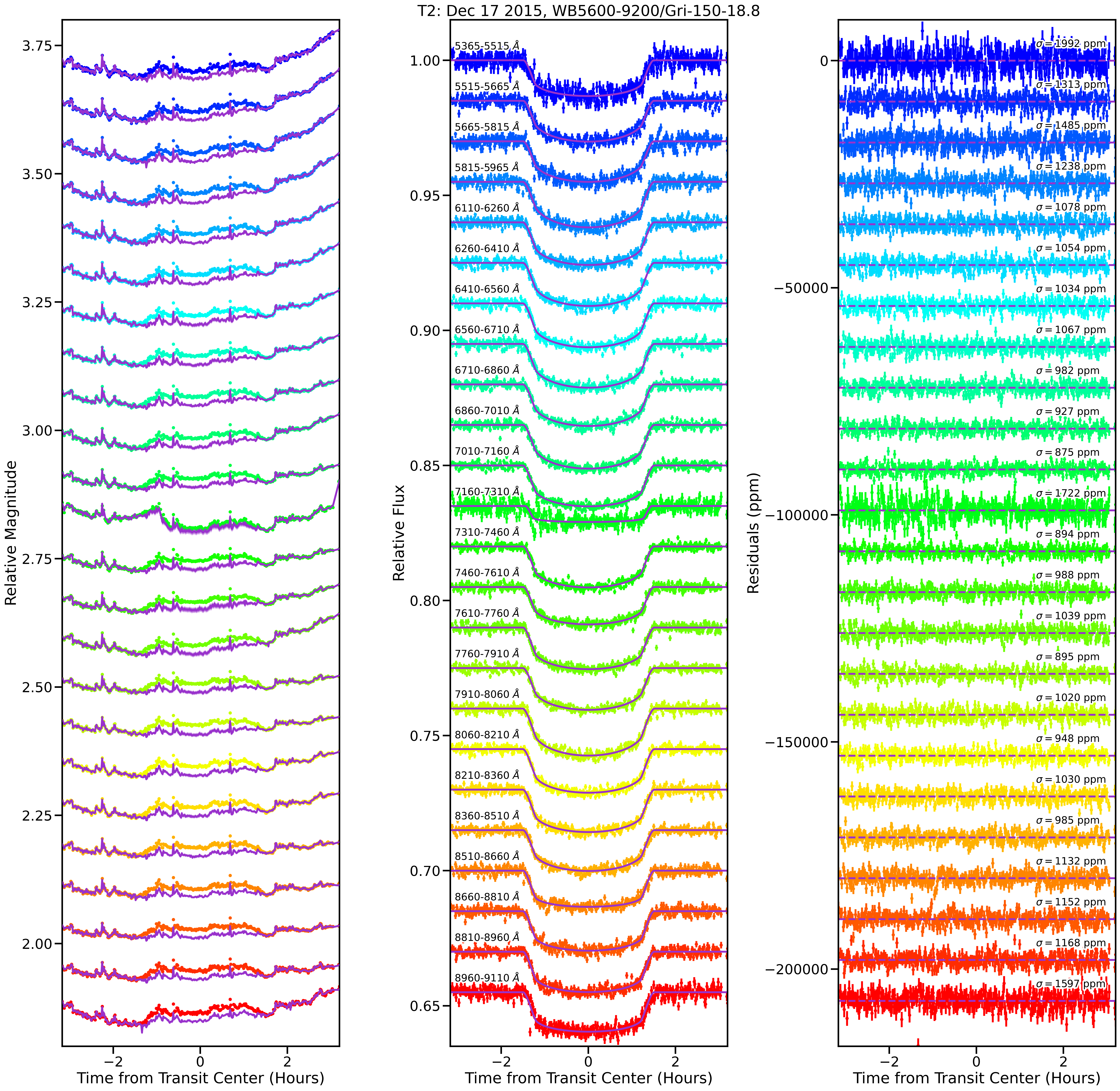}
    \caption{The same as \autoref{fig:wavelength}, but for T2. Note the appearance of Bin 12 (7160-7310 \AA{}), which shows strange behavior and is left out of the final analysis.}
    \label{fig:wlT2}
\end{figure*}

\begin{figure*}[p]
    \centering
    \includegraphics[width = \textwidth]{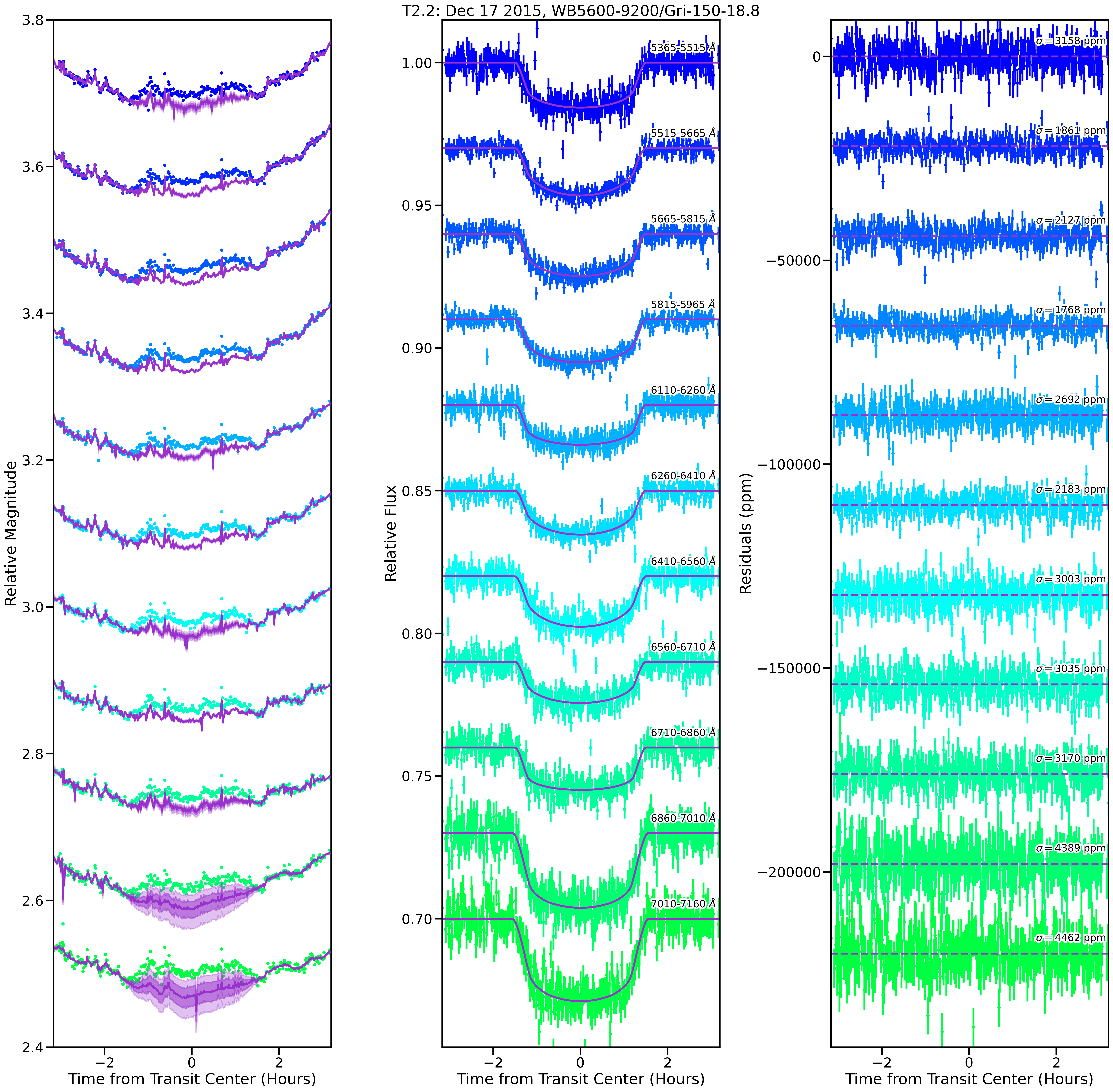}
    \caption{The same as \autoref{fig:wavelength}, but for T2.2. Note that this only covers the bluest 11 wavelength bins, same as is shown in \autoref{tab:bins}.}
    \label{fig:wlT22}
\end{figure*}

\begin{figure*}[p]
    \centering
    \includegraphics[width = \textwidth]{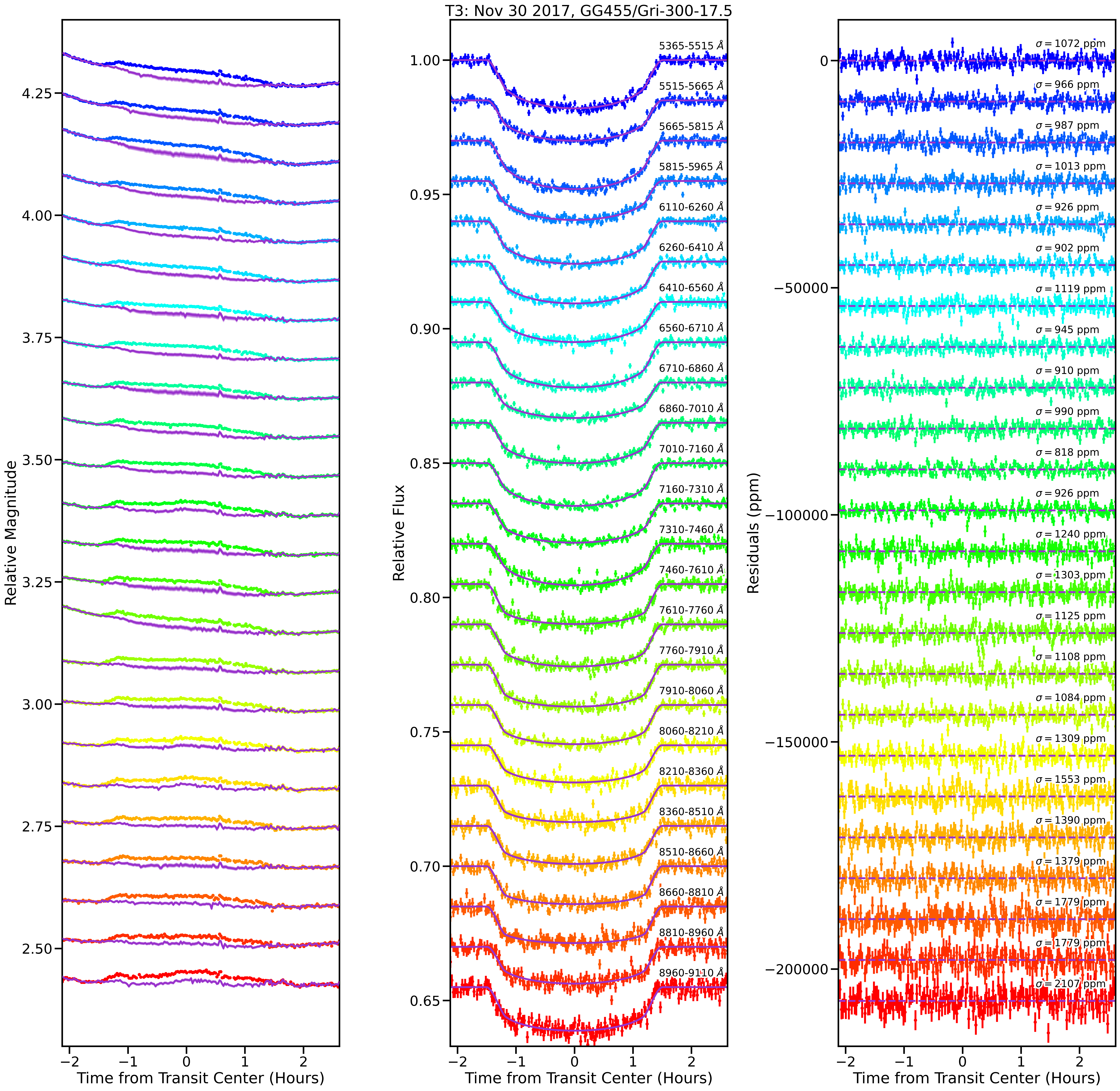}
    \caption{The same as \autoref{fig:wavelength}, but for T3.}
    \label{fig:wlT3}
\end{figure*}

\begin{figure*}[p]
    \centering
    \includegraphics[width = \textwidth]{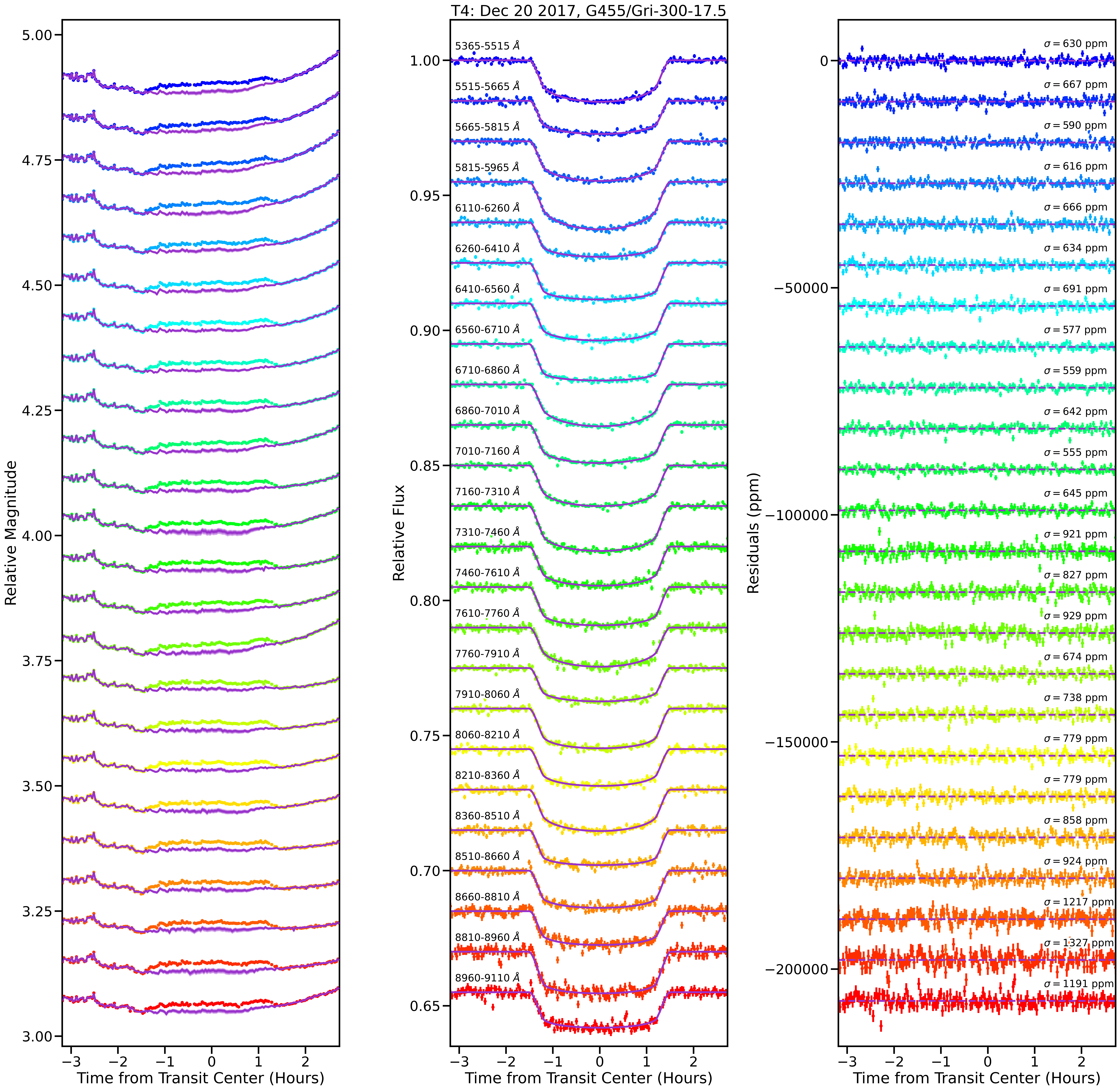}
    \caption{The same as \autoref{fig:wavelength}, but for T4.}
    \label{fig:wlT4}
\end{figure*}

\autoref{fig:wlT1}-\ref{fig:wlT4} show the wavelength-dependent light curves for our transits, besides T5 which is shown in \autoref{fig:wavelength}. 

\pagebreak
\section{Retrievals}\label{app:ret}
\begin{figure*}[h!]
    \centering
    \includegraphics[width=0.9\textwidth]{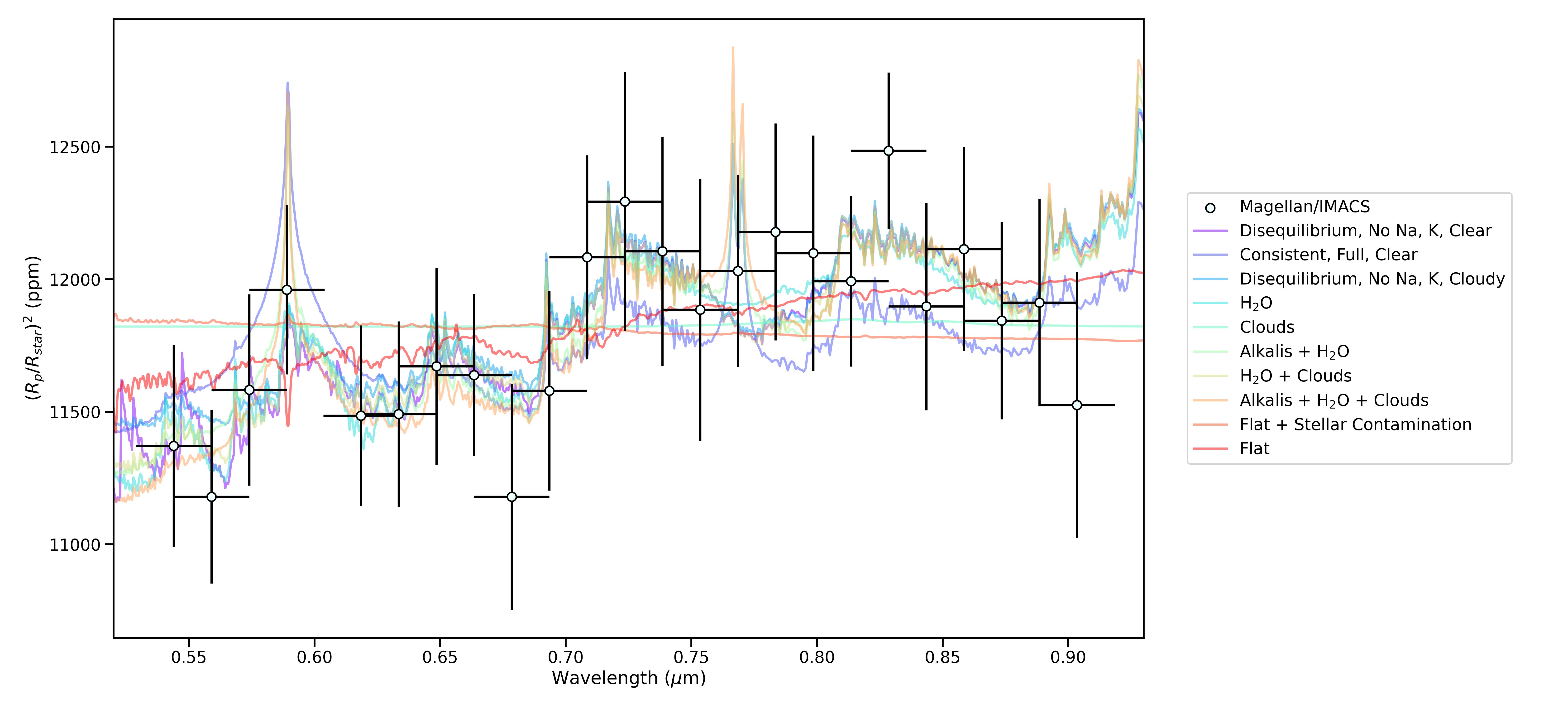}
    \caption{All of the atmospheric models that we tested against our transmission spectrum. The Flat and Flat + Stellar Contamination models are from \textit{Exoretrievals}, and all the rest are from \textit{petitRADTRANS}. Their corresponding $\ln Z$ values are given in \autoref{tab:ret}. }
    \label{fig:retrieval_full}
\end{figure*}

\begin{figure*}[p]
    \centering
    \includegraphics[width=0.7\textwidth]{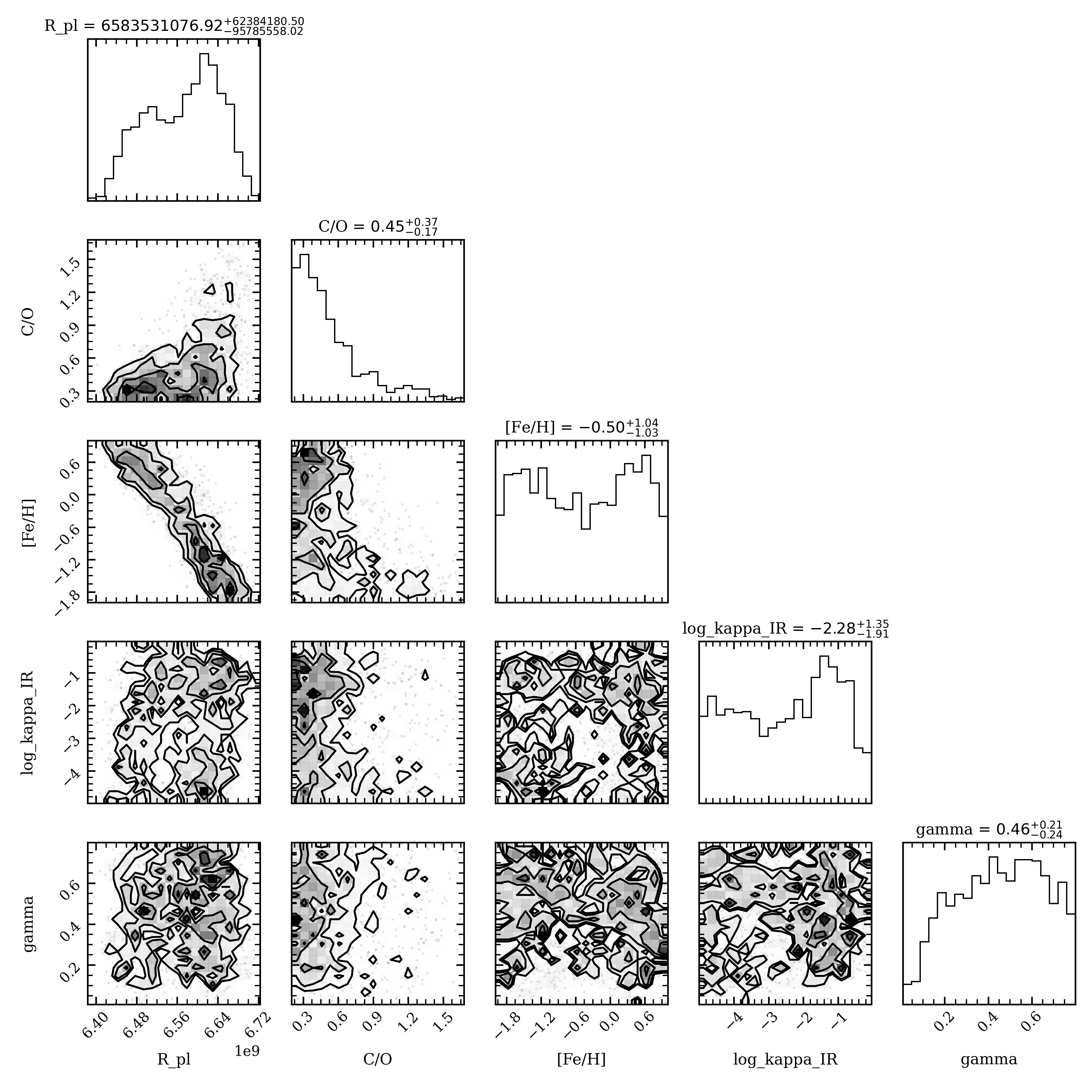}
    \caption{Corner plot showing the posteriors and retrieved best-fit parameters for our best chemically consistent fit, labeled ``Disequilibrium, No Na, K, Clear" in \autoref{tab:ret} and \autoref{tab:retparam}, and in \autoref{fig:retrieval_full}.}
    \label{fig:corner_consistent}
\end{figure*}

\begin{figure*}[p]
    \centering
    \includegraphics[width=0.7\textwidth]{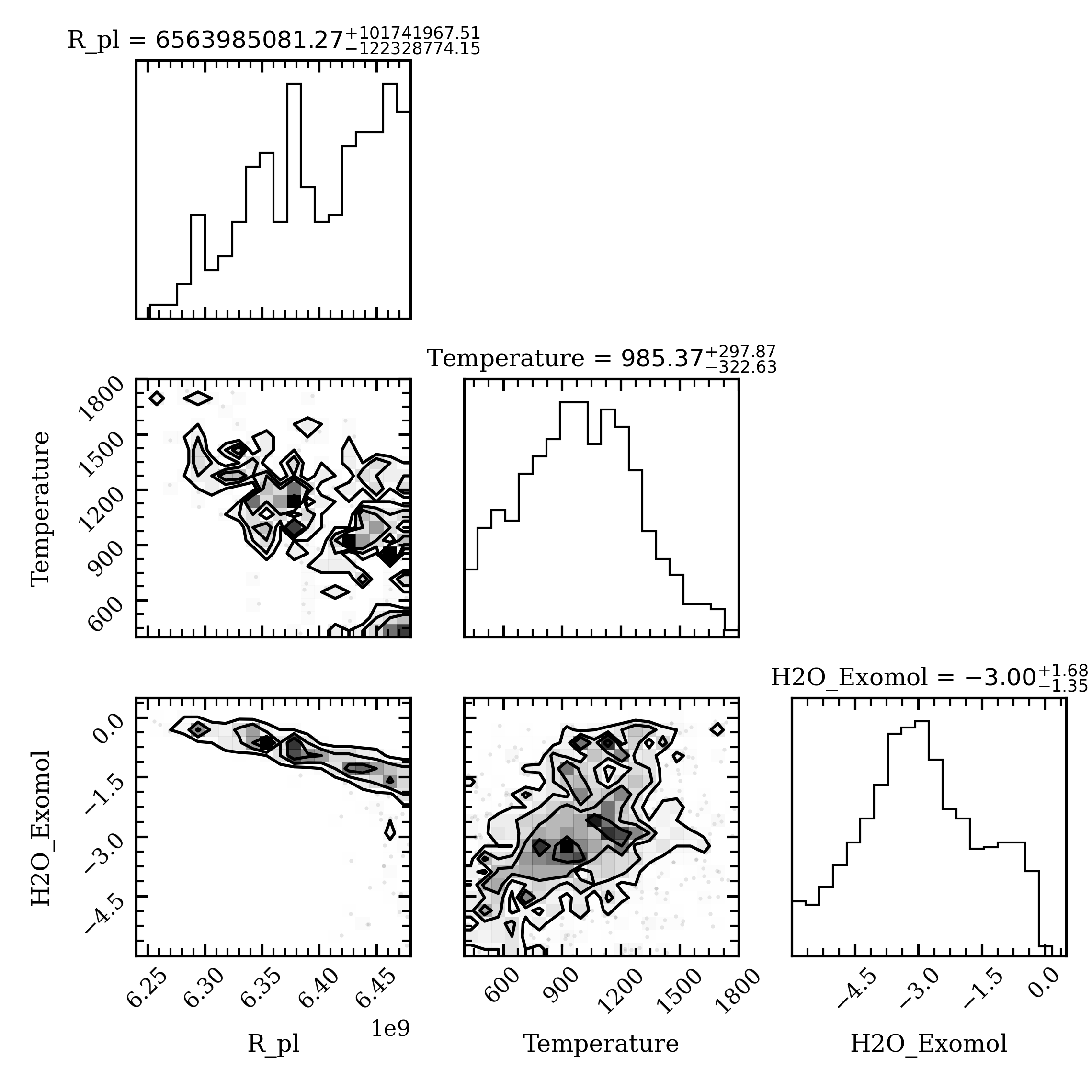}
    \caption{Corner plot showing the posteriors and retrieved best-fit parameters for our best free chemistry fit, labeled ``H$_2$O" in \autoref{tab:ret} and \autoref{tab:retparam}, and in \autoref{fig:retrieval_full}.}
    \label{fig:corner_free}
\end{figure*}

\autoref{fig:retrieval_full} shows all of our tested retrieval models plotted against the final spectrum, and are labeled as given in \autoref{tab:ret}. \autoref{fig:corner_consistent} and \autoref{fig:corner_free} show our retrieval posteriors for our best fit models, ``Disequilibrium, No Na, K, Clear" and ``H$_2$O" respectively.

\pagebreak

\section{Transit-Timing Variations}
Due to the large number of transits covered by our analysis, we additionally look for transit-timing variations in our fit $t_0$. However, we see no evidence for variation from prediction. 
\begin{figure*}[h!]
    \centering
    \includegraphics[width = 0.8\textwidth]{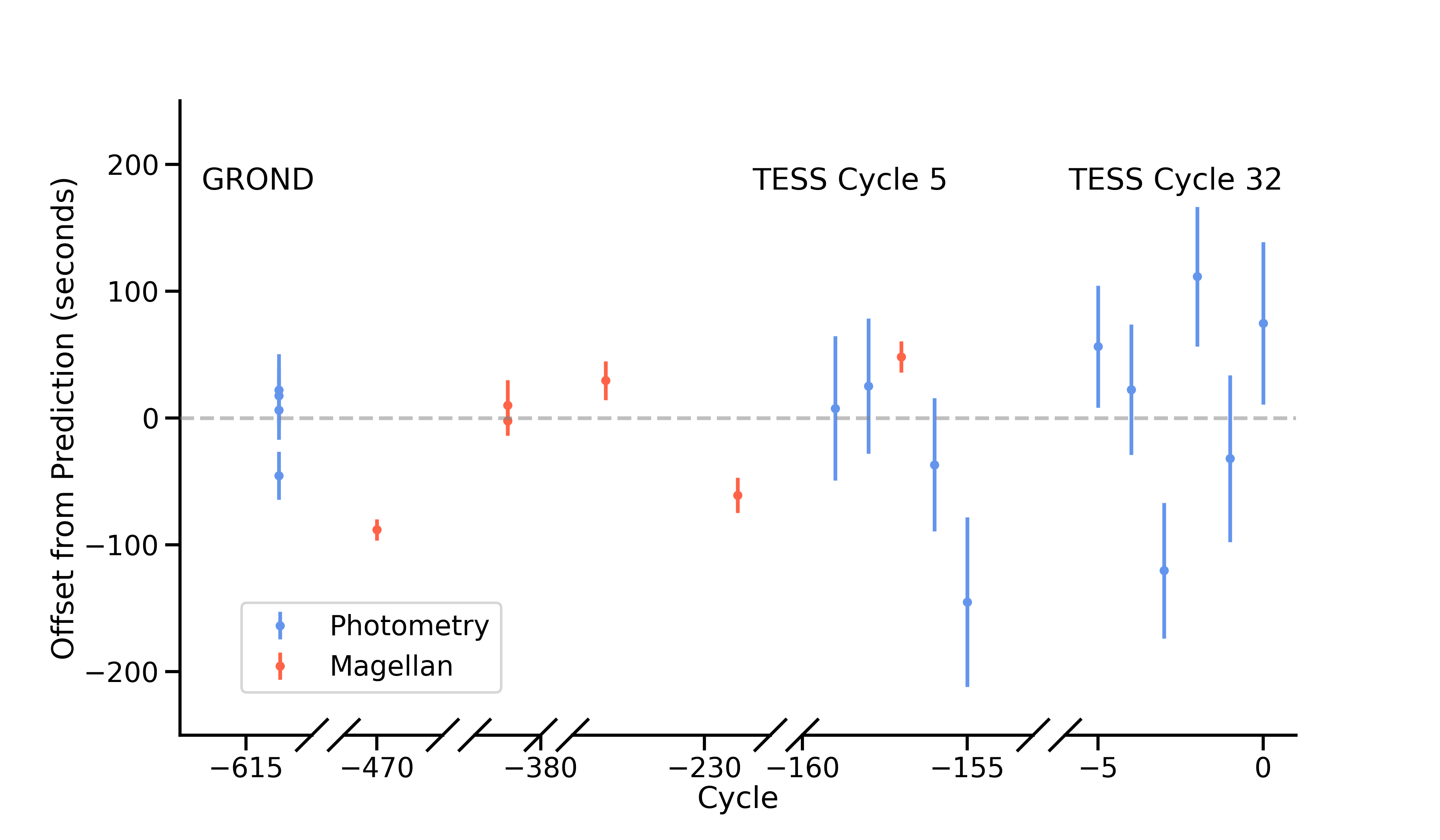}
    \caption{The difference between the predicted $t_0$, found using the $t_0$ and P from \autoref{tab:wlc}, and the fit $t_0$ for both the photometric transits used in \autoref{param_updates} (blue) and our own Magellan transits (red) as are given in \autoref{tab:wlc}. We see no evidence for transit-timing variations in this data. }
    \label{fig:ttvs}
\end{figure*}


\end{document}